\documentclass[aac,crcready]{iosart2x}

\pubyear{0000}
\volume{0}
\firstpage{1}
\lastpage{1}

\usepackage{mathptmx}
\usepackage{soul}\setuldepth{article}
\usepackage{booktabs}
\usepackage{balance} 
\usepackage{xcolor}
\usepackage{multirow}
\usepackage{amsfonts}
\usepackage{amssymb}
\usepackage{threeparttable}

\def\hb{\hbox to 11.5 cm{}}
\newtheorem{definition}{Definition}

\mathchardef\mhyphen="2D

\begin{document}

\begin{frontmatter}

\title{An Explanation-oriented Inquiry Dialogue Game for Expert Collaborative Recommendations}
\runtitle{
Dialogue Game for Expert Collaborative Recommendations}


\begin{aug}
\author[A,C]{\inits{Q.}\fnms{Qurat-ul-ain} \snm{Shaheen}\ead[label=e1]{qurat@iiia.csic.es} \ead[label=e4]{qurat-ul-ain.shaheen@scch.at}%
\thanks{Corresponding Author: Qurat-ul-ain Shaheen, Software Comptence Centre Hagenberg, Hagenberg, Austria; E-mail: quratulainshaheen@gmail.com}}
\author[B]{\inits{K. B.}\fnms{Katarzyna} \snm{Budzynska}\ead[label=e2]{Katarzyna.Budzynska@pw.edu.pl}}
\author[A]{\inits{C. S.}\fnms{Carles} \snm{Sierra}\ead[label=e3]{sierra@iiia.csic.es}}
\address[A]{IIIA, \orgname{CSIC},
Barcelona, \cny{Spain}\printead[presep={\\}]{e1,e3}}
\address[B]{Laboratory of The New Ethos, Faculty of Administration and Social Sciences, \orgname{Warsaw University of Technology},
Warsaw, \cny{Poland}\printead[presep={\\}]{e2}}
\address[C]{\orgname{Software Competence Centre Hagenberg},
Hagenberg, \cny{Austria}\printead[presep={\\}]{e4}}
\end{aug}

\begin{abstract}
This work presents a requirement analysis for collaborative dialogues among medical experts and an inquiry dialogue game based on this analysis for incorporating explainability into multiagent system design. The game allows experts with different knowledge bases to collaboratively make recommendations while generating rich traces of the reasoning process through combining explanation-based illocutionary forces in an inquiry dialogue. The dialogue game was implemented as a prototype web-application and evaluated against the specification through a formative user study. The user study confirms that the dialogue game meets the needs for collaboration among medical experts. It also provides insights on the real-life value of dialogue-based communication tools for the medical community.
\end{abstract}


\end{frontmatter}


\section{Introduction \label{sec:intro}}

As the human society has become more digitally connected, it has developed an increased appreciation for interdisciplinary collaboration \cite{Hennessy2011}. Healthcare is one such domain which has a long tradition of interdisciplinary collaboration amongst different medical experts \cite{Healthcare}. Imagine a distributed health recommendation system where different experts come together to find the best possible diagnosis for a patient. These experts could be human agents, artificial agents or a combination of human and artificial agents. The goal of the collaboration would be to integrate multiple perspectives through knowledge transfer and conflict resolution in order to recommend the best possible diagnosis.  Additionally, the system should offer explanations for its recommendation in order to build trust between the users and the system \cite{Naveed2018}.

As an example scenario, consider the dialogue given in Table \ref{tab:introeg}  between three medical experts represented by  $\alpha, \beta$ and $\gamma$. The first column of Table \ref{tab:introeg} shows the identifier for the statement while the second column indicates the expert name followed by the statement they are making.  The experts are participating in a semi-structured formal discussion similar to the ones that take place in oncology \cite{10vMDT}. They are already aware of the objectives and format of the meeting before the dialogue starts so there is no build up on the objectives of the discussion as is the case in oncology meetings. We cover this in detail in Section \ref{Reqs}. In this example, the meeting is not about a cancer diagnosis, but rather a more general diagnosis for a patient denoted by $A$. This is a contrived example, aimed at illustrating the conversation flow among healthcare experts who are participating in a semi-structured discussion to collaboratively make a recommendation for a patient. The goal of this example is not to provide a comprehensive medical discussion on the patient's diagnosis as this is not the objective of this work. Moreover, note that the dialogue only shows the knowledge and reasoning of the experts to the extent that they choose to reveal through their conversation. It does not show the complete knowledge base or the reasoning process of the expert agents (who can be human or artificial). 

\begin{table*} [t] 
  \caption{Example dialogue between three medical experts $\alpha, \beta, \gamma$}
  \label{tab:introeg}
  \begin{tabular}{p{0.04\textwidth}p{0.90\textwidth}}\toprule
    \textit{Id} &\textit{Dialogue}  \\ \midrule
	1 & $\alpha$: Patient A is a 48 years old female. Her symptoms are fatigue, constipation, weight gain, drowsiness, and dry skin. It looks like a case of depression. I recommend lifestyle changes such as healthy eating and walking. I note that she has a family history of autoimmune diseases. \\
	2 & $\beta$: Can you justify your diagnosis of depression?\\
 	3 & $\gamma$: I would say that it looks like a typical case of hypothyroidism.  \\
	4 & $\alpha$: Because her symptoms are typical of depression.\\
	5 & $\gamma$: I note that she does not have headache or back pain, which are common in case of depression. \\
	6 & $\beta$: Can you justify why you diagnose hypothyroidism?\\
	7 & $\gamma$: Because she has dry skin and a family history of auto immune diseases, which are typical for hypothyroidism. \\
    8 & $\alpha$: You might be right. I recommend  that we test her TSH, T4 and T3 levels. \\
    9 & $\beta$: I agree. I also recommend doing a blood complete count to rule out other similar conditions like anaemia. \\
    10 & $\gamma$: I agree to testing TSH, T4 and blood complete count. I think we can close the discussion now. \\
    11 & $\beta$: I agree. \\
    12 & $\alpha$: I disagree. What about testing for T3? \\
    13 & $\gamma$: Why do you want to test for T3? \\
    14 & $\alpha$: Because I want to rule out Hyperthyroidism. \\
    15 & $\beta$: Yes, it makes sense. \\
    16 & $\gamma$: I don't think it is necessary to test T3 at this stage since she is not asymptomatic. And her symptoms are closer to Hypothyroidism. \\
    17 & $\alpha$: Okay. I think we can close the discussion now. \\
    18 & $\beta$: I agree. \\
    19 & $\gamma$: I agree. \\
   \\ \bottomrule
  \end{tabular}
\end{table*}

 In this example, we assume that agent $\alpha$ loosely represents a clinical psychologist, agent $\beta$ a general practitioner and agent $\gamma$ an endocrinologist.  Agent $\alpha$ first presents the facts of the case and offers his own diagnosis (depression). This is challenged by agent $\beta$. $\alpha$ then justifies their stance which is rejected by $\gamma$. $\gamma$ then proposes their own diagnosis. $\beta$ asks $\gamma$ to explain their diagnosis. Subsequently, both $\alpha$ and $\beta$ accept $\gamma$'s explanation. Then $\alpha$ proposes some tests to which $\beta$ agrees and propose an additional test. $\gamma$ agrees and proposes yet another test. However, their suggestion is ignored by the other two as they propose to close the discussion. However, $\gamma$ does not consent and asks the other agents to respond to their suggestion. They do respond and subsequently $\gamma$ agrees to end the discussion. Throughout the discussion, the agents try to collaborate in a cooperative manner. We use this dialogue as a running example throughout the rest of the paper to illustrate our approach.  

\begin{figure}
	\centering	\includegraphics[width=0.75\textwidth]{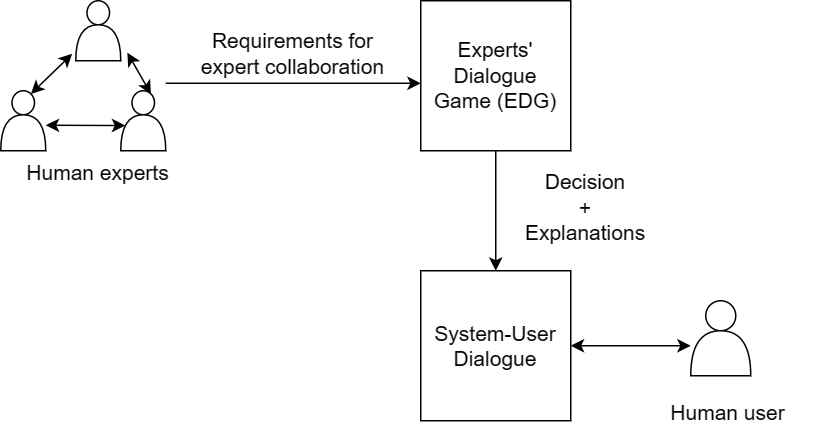}
	\caption{Workflow diagram of an explainable multiagent recommendation system employing EDG.}
	\label{motiv}
\end{figure}

As a first step towards realising such a hybrid human-artificial multiagent system capable of such a dialogue, we propose a novel interaction protocol between experts agents (whether human or artificial). We call this protocol as \emph{Experts' Dialogue Game (EDG)}. Figure \ref{motiv} visualises how EDG would fit into the pipeline for building such a system. As a first step, real-life consultations among medical experts are formalised as a requirement specification. Building on this specification, EDG is defined as a dialogue game among the participants in a hybrid human-artificial multiagent system (MAS).  The output of EDG is a recommendation from the system along with a sequence of explanations justifying the recommendation. These explanations can then be plugged into a system-user dialogue to justify the system's recommendation to the user.  The user in this case could be the patient himself or a physician in charge of the patient.  While most of the current literature \cite{Walton2011,Arioua2017EXplain,Madumal2019,Dennis2021,Ilia2023} focuses on explaining the working of a system to a human user through a system-user dialogue, EDG investigates how an explanation dialogue can be used within a multiagent system as part of agent reasoning.

 Dialogue games are \emph{dialectical systems} \cite{Hamblin1970} in the tradition of informal logic \cite{Walton1995} and are formally defined as verbal interactions between two or more players according to some pre-defined rules for the dialogue \cite{Iyad2007}. Each interaction is specified with the use of \emph{locutions} which represent speech acts permitted in a given dialogue \cite{Searle1969}. Dialogue games require that each participant maintains consistency across its statements, also called \emph{commitments}, at any point in the dialogue \cite{Walton1995}. Characteristics of these verbal interactions are typically defined  in multiagent communication  according to the popular typology introduced by Walton and Krabbe \cite{Walton1995}. Dialogue games have a long tradition of being used to solve formal problems as well as to model natural language communication in real-life settings. In the first case, they have been employed to search for formal logical proofs in the tradition of Lorenzen \cite{zora2020} and leading to the field of \emph{dialogical logic}, as well as in the \emph{prescriptive} approach such as Hamblin's system \cite{Hamblin1970}, to disallow logical fallacies during natural language argumentation. In the second case, they are used to study the communication dynamics in real life settings, giving rise to the \emph{descriptive} approach to dialogue games \cite{PrakkenLaw,purchaseNego2003,Madumal2019,Snaith2016,Snaith2018,Olena2014}. This approach can also be used to inform computational models of interactions between agents in a multiagent system.  

The goal of EDG is for the participants to collaboratively find the best recommendation through exchange of knowledge and mutual agreement. In Walton-Krabbe's typology, this scenario fits an \textit{inquiry} type of dialogue in which an initial situation is a need to have proof (i.e. in our case --- to find the best recommendation), the participants' goal is to find and verify an evidence  (i.e. to consult an observation), and the goal of the dialogue is to prove or disprove a hypothesis (i.e. to argue for or against a recommendation). This is in contrast to a \emph{deliberation dialogue} where the initial situation is a need for action, the participants' goal is to influence the outcome and the goal of the dialogue is to reach a decision on the best possible action \cite{Walton1995}. In order to generate explanations of the recommendations, we define agents' communicative behaviour in the dialogue through \textit{explanation-based} illocutionary forces which can then be traced back and retrieved in response to a query and presented to the user as an explanation of the recommendation: (1) \textit{explanation requests} such as $wh\mhyphen explain (p)$ when the speaker knows that $p$ is the case, but does not understand why it is the case; $\mathit{wh\mhyphen justify} (p)$ when the speaker does not agree that $p$ is the case and asks the hearer for the justification of $p$; and $\mathit{wh\mhyphen clarify} (p)$ when the speaker does not understand a term in  $p$ and asks for the clarification of this term; and  (2)\textit{ explanation replies} such as $explain (p)$ when the speaker provides an explanation for $p$; $\mathit{justify} (p)$ when they provide a justification of $p$; and $\mathit{clarify} (p)$ when they provide a clarification of the term in $p$.

The contribution of this work is threefold. First, we present a requirement specification for collaboration between experts. While the requirement specification is grounded in the medical domain, it focuses on the general communication issues that can come up during expert collaborations. Hence, it can be abstracted to consultations among experts in general.  Next, inspired by the tradition of dialogue embedding \cite{Walton1995}, we combine explanation-based illocutionary forces in an inquiry dialogue to generate richer traces for the inquiry process than what is possible with the \emph{assert} locution,  typically used in inquiry dialogues. Moreover, the dialogue game is unique in that sense that it meets the requirement specification from the domain experts. While the combination of different dialogue types is not ground breaking formally, the proposal makes an important methodological step into the applications of such formal dialogue systems into real-life domains which is empirically grounded in users' requirements.  Furthermore, to the best of our knowledge, no other \emph{descriptive} dialogue game has focused on interactions among experts.  The rich game traces allow the system to be transparent and can be used to explain the recommendations of the system to a human user through human-machine interfaces such as verbal and visual interaction. Finally, we evaluate the dialogue game against the requirement specification and verify the evaluation through a formative user study. Thus, we introduce the methodology of user-centred software engineering practice to dialogue games. While other works \cite{PrakkenLaw,purchaseNego2003,Madumal2019,Snaith2016,Snaith2018,Olena2014} in \emph{descriptive} dialogue games have used insights from their domain of interest to inform the dialogue games, none of them have provided a requirement specification as far as we are aware.

The rest of this work is structured as follows. Section \ref{Lit} summarises related work. Section \ref{Reqs} presents the requirement specification for expert collaborations. Section \ref{game} formally presents the Experts' Dialogue Game (EDG) while Section \ref{platform} provides implementation details of a web-based platform implementing the EDG. In Section \ref{eval}, EDG is evaluated against the requirement specification presented in Section \ref{Reqs}. Section \ref{insights} highlights user perspectives on EDG and the platform described in Section \ref{platform}. Finally, Section \ref{Conc} concludes the paper and provides directions for future work.

\section{Related Work \label{Lit}}
This section summarises related work on explanatory and inquiry dialogues, dialogue games in multiagent systems for healthcare, argumentation in healthcare and communication tools for multidisciplinary collaborations in healthcare. 

Amgoud et al. \cite{Amgoud2000} present an argumentation system for resolving inconsistencies in an agent's knowledge base. Subsequently, they show how dialogue game theory can be applied on top of this to realise the different dialogue types in Walton and Krabbe's typology \cite{Walton1995}. A minimal framework for an explanatory dialogue system is presented in Walton \cite{Walton2011}. The goal of the dialogue is for an \emph{explainer} (an entity that explains, usually the system) to fill in the gaps in the knowledge base of the \emph{explainee} (the target of the explanation, usually a human user) by informing them why something happened. This is considered as a transfer of understanding from the explainer to the explainee.  Building on Walton's minimal framework, Arioua et al. \cite{Arioua2017EXplain} combine an explanatory dialogue with argumentative illocutions. The explanatory illocutionary force is used  by the system to explain the behaviour of some phenomena to the user while the argumentative force helps to resolve inconsistency in the knowledge bases of participants. In addition to commitment stores for each participant, they introduce an \emph{understanding store} for the user which stores the missing links in understanding rather than what is currently understood. The discharging of all issues in the understanding store confirms a successful transfer of understanding. 

Madumal et al. \cite{Madumal2019} analyse human-human and human-agent explanatory dialogues from various domains and propose an explanatory dialogue protocol based on induction. Their protocol also combines an explanatory dialogue with argumentative faculty. The goal of the dialogue is the same as proposed by Walton \cite{Walton2011}. They use double acknowledgement for confirming understanding, that is, acknowledgement by the explainee on being satisfied with the explanation and explainer's return acknowledgement. Dennis et al.  present an explanatory dialogue game for explaining the behaviour of a Belief-Desire-Intention System \cite{Dennis2021}. Their protocol, like EDG, is grounded in dialogue game theory from informal logic using argumentation schemes and critical questions rather than argumentation theoretic dialogue \cite{Walton2009}. The goal of the dialogue in this case is to understand system behaviour through comparing traces of the system from different participants. Similar to these works, we use explanation-based illocutionary forces to fill in missing gaps in the  knowledge bases of participants.

Ilia et al. \cite{Ilia2023} extend an information seeking dialogue with explanatory illocutionary forces. The goal of the dialogue is to offer factual and counter factual explanations of a classifier's output to a human user. They also evaluate the protocol through a user study and process analytics. Prakken and Ratsma \cite{RatsmaPRakken2022} propose a \emph{case-based explanation dialogue} to explain the outcome of a linear binary classifier. The goal of the dialogue is to provide a model-agnostic local explanation. The explanations in this case are not trying to explain the system reasoning but rather trying to come up with reasons to justify the system result. The dialogue game starts with the proponent of the dialogue presenting a similar example from the training set with the same outcome as the current instance. The opponent can argue against this using two strategies. The first is to use counter examples from the training set. The second is to highlight the differences between the current instance and the instance being presented as a \emph{justification} by the proponent. A successful explanation amounts to a winning strategy for the proponent. However, unlike these approaches, we incorporate the explanatory illocutionary forces in the reasoning process itself rather than to only explain why the system behaved in a certain way. While these works target a transfer of understanding from the system to a human user, EDG involves explanations amongst the reasoning agents. 

Although not as common as persuasion dialogues, a few other works have explored inquiry dialogues as stand alone dialogues or in combination with other dialogue types. 
Bex et al. \cite{Bex2008} combine an inquiry dialogue with a persuasion dialogue for discussions between criminal investigators. The goal of the dialogue is to come up with the most robust explanation. They assume an adversarial setting in which each agent advocates for its own preferred explanation. Unlike this work, EDG assumes a cooperative setting where the main goal is to come up not only with the most robust explanations but decisions as well. Black and Atkinson propose a framework that embeds an inquiry dialogue over beliefs with a persuasion dialogue over actions. The inquiry dialogue allows the participants to collaboratively decide what to believe whereas the goal of the persuasion dialogue is to collaboratively decide what is the best action to do in order to reach the proponent's goal. Once all the arguments have been given, it is upto to the proponent of the dialogue to make the final decision based his personal preferences \cite{Black2009}.

Black and Hunter \cite{ArgInquiry2007, WarrantInquiry2009} propose two types of inquiry dialogues, which they call as \emph{argument inquiry} and \emph{warrant inquiry}. The goal of \emph{argument inquiry dialogue} \cite{ArgInquiry2007} is for two agents to jointly construct an argument for a claim by sharing relevant beliefs. The protocol has three moves and allows nesting of argument inquiry dialogues. In addition to a conventional commitment store, they also introduce a \emph{question store} which keep tracks of the premises that need to be proven in order to prove the claim representing the dialogue topic. They prove soundness and completeness for their protocol. The goal for a \emph{warrant inquiry dialogue} \cite{WarrantInquiry2009} is for two agents to share arguments to jointly construct a \emph{dialectical tree} in order to determine the acceptability of a particular argument. The main difference between the two types of dialogues is that argument inquiry is not concerned about the acceptability of the argument constructed while the latter is. Warrant inquiry dialogue allows embedding argument inquiry dialogue and also involves a question store like the former. They also provide a strategy for selecting the next move for a participant for both types of dialogues. However, to the best of our knowledge, none of these works combine an explanatory dialogue with an inquiry dialogue to make agent reasoning explainable.

Other works have incorporated argumentation and dialogue games in multiagent systems to provide clinical decision support in a distributed environment such as cancer diagnosis and management. Huang, Jennings and Fox \cite{Huang1995,Huang1994} present a multiagent architecture for medical decision support in an interdisciplinary setting. The architecture has four components; a three layered knowledge base, a centralised working memory, a communications manager and a human-computer interface. The architecture also covers decision making under uncertainty, task management and agent cooperation. The communication manager works with communication \emph{primitives} or locutions and a communication protocol. However, the locutions in this case are geared towards managing the tasks in a distributed environment rather than a discussion amongst different experts. For example, the locutions \emph{request, accept, reject} and \emph{alter} are used in the task allocation stage. The locution \emph{inform} is used to report on the allocated task while the locution \emph{propose} is used to recommend a treatment plan in response to a \emph{query}. In contrast, EDG is focused on facilitating the discussion amongst the experts rather than a distributed management of responsibilities.

Beveridge and Fox \cite{Beveridge2006} use a dialogue game as an interface between the underlying task structure and ontological knowledge and the spoken dialogue generation system. They implement their approach to provide clinical decision support for breast cancer diagnosis. They use several locutions as \emph{initiating} locutions. For example, \emph{inform} is used to present new information, \emph{instruct} is used to request an action from the user, \emph{$query\mhyphen yn$} to ask a question with a yes or no answer and \emph{$query\mhyphen w$} to elicit a value from the user as part of data entry. They treat the dialogue started from each initiating locution as a sub-game. In contrast, the query locutions in EDG are targeting towards incorporating different types of explanations into the multidisciplinary discussion amongst experts. This is because the goal for EDG is to support collaborative discussion among experts rather than supporting an individual user with decision support.

Vasileiou et al. \cite{Vasileiou2023} present an argumentation-based justification dialogue between two participants. The explainee is a human user who wants to understand the explainer's (artificial intelligent agent) reasoning. The dialogue game has four locutions, two of which are reserved for each of the participants. The game only allows a single locution per turn. They provide evaluation of their dialogue game through a user study and discuss its properties. Rago et al. \cite{Rago2023} present the notion of explanatory dialogue between two participants as an \emph{Argument eXchange (AX)}. They discuss desirable properties of AX for agents equipped with quantitative bipolar argumentation frameworks and gradual semantics. \cite{Sassoon2019,Sassoon2021,Kokciyan2021} propose an interactive clinical decision support system, called CONSULT, for multimorbidity patients to self-manage their treatment. The system integrates four types of data sources; the patient's electronic health record, data from sensors monitoring the patient's symptoms, the clinician's input and finally treatment guidelines. It uses computational argumentation to aggregate the data and resolve inconsistencies in the data sources. It also provides an argumentation-based dialogue interface for system-patient interaction to interactively deliver the recommendations to the user. The system-patient interaction uses templates for its text-based natural language interface. It is based on three different argumentation schemes and their associated critical questions. These cover deliberation, persuasion and explanation dialogues. Castagna et al. \cite{EQRbot} propose an explanation dialogue between two participants that also interfaces with the CONSULT system through a chatbot. They propose an argument scheme based on practical reasoning and use it for the explanatory dialogue. Shaheen et al. \cite{Shaheen2021} propose an explanatory dialogue between two participants to explain the recommended treatment plan for multimorbid patients.  All of these works focus on explanation dialogues between two participants, mainly a system as an explainer and a human as an explainee.  In contrast to these works, EDG aims to use different types of explanatory dialogue forces to generate richer traces during inter-agent reasoning processes.

Pancho et al. \cite{Tolchinsky2006,Modgil2005Pnacho} propose an argumentation-based \emph{deliberation} dialogue between two agents to discuss the viability of transplant organs. The dialogue is implemented as part of \emph{Carrel+}, a health information system to manage organ transplants in Spain. The dialogue model, called \emph{ProCLAIM}, \cite{Tolchinsky2007} is based on argument schemes and case-based reasoning. ProCLAIM employs a mediator agent to guide the participant agents on their legal moves, decide the validity of submitted arguments and finalise the recommendation regarding the viability of the proposed transplant organ. The mediator agent uses argument schemes, existing guidelines, case-based reasoning and a component manager to manage the strengths of submitted arguments. It applies abstract argumentation semantics \cite{Dung1995} to decide the winning argument. 

Xiao et al. \cite{Xiao2021} present a group decision description language and a consensus protocol for a multiagent system. However, the protocol is not based on speech act theory \cite{Searle1969}, but is an agent communication protocol that uses functions like averaging and intersection to generate consensus values unlike the work presented here. Patkar et al. \cite{Patkar2012} developed a clinical decision support tool, called \emph{MATE (Multidisciplinary meeting Assistant and Treatment sElector)}, to support multidisciplinary cancer conferences. The tool is responsible for information management and providing treatment recommendations after processing the data. It does not present a dialogue game to support multidisciplinary discussion amongst experts as is presented here. A comprehensive survey describing the use computational argumentation for explainable artificial intelligence can be found in Vassiliades et al.\cite{ArgXAISurvey2021}. Some other works have proposed computational argumentation systems for clinical decision support \cite{FToniKR12,FanToni2013}. \cite{} present a negotiation protocol for agents in a Belief Desire Intention (BDI) architecture. However, the protocol is grounded in agent communication language rather than dialogue game theory. 

The term Health Information Technology (HIT) refers to the application of information technology to facilitate healthcare. HIT systems fall on a wide spectrum ranging from administrative support, patient information management and retrieval, communication and decision support \cite{Chaudhry2006}. Carayon et al. \cite{Carayon2019} point out that most of the existing HIT systems are focused towards individual tasks rather than teams, even as team-based care is becoming becoming a popular paradigm. Here we review representative HIT applications targeting multidisciplinary communication and collaboration support. 

\emph{Care Connector}  \cite{Tang2018,Tang2019,Tang2023} is a communication and collaboration platform  implemented in a community hospital in Canada. It is a web application that integrates into the HIT of the hospital to retrieve and update electronic records. The application covers both care planning and monitoring modules in addition to a messaging module between multidisciplinary care providers. The patient information is stored as part of a \emph{Care Planner} module. The messaging modules provides asynchronous communication of non-urgent messages using the information in the Care Planner as a shared knowledge base. The messaging module allows linking each conversation to a patient. It informs participants to post messages following the Situation, Background, Assessment and Recommendation (SBAR) framework which is used in healthcare communication. However, it does not force the participants to frame their messages according to this framework.

Kurahashi et al. \cite{Kurahashi2018} present another communication and collaboration tool called \emph{Loop}. It allows multidisciplinary collaboration between teams. The teams can include healthcare professionals, caregivers as well as the patient. Each conversation loop is centred around a patient. The application includes a card with patient information on the left hand side while the right hand side has the messaging thread. The information exchange is secure and sharing information between different subgroups is allowed. For example, a professional only message exchange or with all the participants in the loop. It also allows tagging messages with user-defined labels to facilitate search later on. The labels represent different themes or \emph{issues} described in the message. 

\cite{Lin2020} implemented a  platform, called \emph{one-stop platform}, for multidisciplinary collaboration among healthcare professionals in a Taiwanese hospital. The platform integrates into existing HIT system of the hospital. It covers administrative and planning aspects in addition to a messaging module. The messaging modules allows transparency and accountability for message posting and viewing. It supports exchanging text, audio and video messages. However, there is no specification on the format of the content that is exchanged. 

\emph{Shared Care Platform (SCP)} \cite{MartinezGarcia2013a} is yet another collaboration tool implemented in a hospital in Spain that builds on social networking and open source tools. The tool is targeted towards facilitating healthcare professionals to manage multimorbidity patients. It has two components; a social networking component, called the \emph{Clinical Wall}, and a decision support component. The Clinical Wall provides social networking like collaboration and communication support amongst healthcare professionals. It is integrated into the electronic health records of the patient. The record has an assessment section, a discussion section and a conclusion section. The assessment section includes information on patient history and assessments. The conversation starts with one clinicians posing a question to others. In the discussion section the clinicians exchange messages to arrive at an agreement with regards to the question. In the conclusion section all participants need to sign off on the agreed decision. During the discussion any clinician can be added to the conversation to invite their feedback. The decision support component uses the Clinical Wall and provides clinical guidelines in the form of rules. Other works \cite{Ngo2020,Morse2021,MobApps2020,Wu2011} implement mobile applications to support care and communication amongst healthcare professionals, caregivers and patients. These applications mainly support administrative and information management tasks with simple messaging support for communication. 

All these platforms and applications provide secure messaging amongst participants and well integrated interfaces for the existing HIT systems in place. In contrast, the prototype implementation provided for EDG does not provide any of these features since the goal in this work was to evaluate the underlying protocol rather than present a full-fledged web application. However, none of the existing applications provide support for framing the content and type of messages with an underlying dialogue protocol as is proposed in this work. So the platform provides a novel idea of supporting collaborative communications amongst healthcare experts based on an underlying dialogue protocol.   

\section{Requirements for Expert Collaboration \label{Reqs}}
In this section, we propose a requirement specification for successful consultations among experts. Consultations among experts are common in the professional world, especially when critical decisions are concerned such as in medicine, aviation and engineering. We focus on consultations among medical experts as our domain of choice in order to develop a dialogue protocol for consultations among expert agents. This is because it is easier to abstract away from domain specific terminology in this case in order to understand the interaction dynamics. However, the requirement specification we present is abstract enough to be applied to consultations among experts in general since it avoids domain specific scenarios and terminology. 

In order to understand collaboration scenarios between experts, we held informal discussion with some medical experts (specifically a gynaecologist, a radiologist, a general physician and a dentist).  We identified two main scenarios, informal consultations such as during hand-off of a patient from an emergency room to a general or specialist ward, and formal consultations which usually take the form of \emph{multidisciplinary cancer conferences} or \emph{case conferences} for short. These conferences are structured discussions between different specialists to finalise diagnosis and treatment options for cancer patients \cite{Wright2007}. The conference is attended by multiple specialists such as surgery, oncology and pathology. It starts off with the specialist in charge presenting each case history to the panel of experts. This is followed by a discussion amongst experts as to the best possible diagnosis and treatment options for each patient \cite{Gross1987}. During the discussion, knowledge transfer between experts takes place. This happens through the \emph{explanatory, inquisitive} and \emph{cooperative} tone of the dialogue. Because of its explanatory value, many specialists consider it to have educational value for trainees \cite{Gross1987}. We chose the case conference as our main use case to inform the protocol because of its formal and structured format. Moreover, some of the general communication issues \cite{Sutcliffe2004,8CommBreakdown,9Barriers} during informal hand-off also come up in case conferences. Subsequently, we reviewed medical literature on communication in case conferences to identify possible issues. We also included some works on general communication issues during informal collaboration between medical experts.  These works were included since they were general enough to be understood by non-medical audiences.

Sutcliffe et al. \cite{Sutcliffe2004} identify two types of communication failures in the medical profession: \emph{systematic} and \emph{individual}. While systematic failures result from a lack of sufficient organisation, individual failures have complex roots such as hierarchical and power dynamics and excessive workload. They suggest establishing communication guidelines to minimise both types of failures. In order to mitigate against these failures, we develop an interaction protocol grounded in the interaction dynamics during case conferences \cite{Delaney2004} as well as general communication dynamics \cite{Sutcliffe2004} that can come up during informal collaboration between medical experts as a result of organisational subculture \cite{Sutcliffe2004}. Formally, we used  Scopus, PubMed and GoogleScholar to look for papers from the medical community that identify communication issues in cancer conferences and in general. We used the keywords `communication multidisciplinary cancer conference', `multidisciplinary cancer conference'  `communication failure medical experts' and `tumour board decision making'. All open-access, English language articles between 2001 and June 2021 related to medicine were considered. Amongst these, manual filtering was done to narrow down results to works involving reflections on communication issues amongst medical experts in cancer conferences and in general. The included articles were either reporting reflections from user studies \cite{GermanStudy,9Barriers,Delaney2004,15MDCobstacles,16Teamwork,12EHR,Sutcliffe2004,8CommBreakdown}, surveys \cite{Macaskill2006,13MDCoppur,Rajasekaran2021,Wright2007} or best practices \cite{Wright2007,10vMDT,12EHR} followed by professionals.  Works related to communication between patients and healthcare professionals were excluded. As were works that focus on the diagnostic recommendations for different medical conditions. We stopped our search for articles when the same ideas started to recur in different articles and we felt confident that new articles were not adding any new perspectives. Eventually, fourteen articles were included in the study which are given in Table \ref{tab:reqs}. 

We identified fourteen basic requirements for consultations among medical experts which are presented in Table \ref{tab:reqs}.  These cover both the \emph{systematic} and \emph{individual} needs for effective consultations between experts. A requirement was considered as inferred from a publication if it was explicitly or implicitly mentioned as a standard practice, a desired outcome or as a lack thereof. 
All the best practices, guidelines, reflections in the papers were taken into account, grouped together and summarised. This robust process of  systematic and rigorous data collection from the domain literature is treated as providing validation of requirements which are then further evaluated in the user study on requirements embedded in the dialogue protocol (see Section 6). These requirements adhere to the structural guidelines and best practices for case conferences while at the same time addressing the communication issues that come up during informal consultations. They are abstract enough to be applicable in a formal collaboration setting between experts in different domains. They can be seen as sub-goals that can facilitate the collaboration in order for it to be productive. 

\begin{table*}[t]
  \caption{Requirements for effective communication between experts according to medical literature}
  \label{tab:reqs}
  \begin{tabular}{rlll}\toprule
    \textit{Id} &\textit{Requirement} & \textit{\#papers} & \textit{References}\\
    & & (Total: 14) & \\ \midrule
    & \textbf{Agent Oriented Requirement, $RA$} &  & \\
    RA1 & To minimise the effects of individual constraints in communication. & 7 & \cite{Sutcliffe2004,Kagan2005,Delaney2004,8CommBreakdown,9Barriers,10vMDT,16Teamwork}\\
    & \textbf{Cooperation Oriented Requirements, $RC$} &  & \\
    RC1 & To enable and promote cooperation. & 7 & \cite{Sutcliffe2004,Rajasekaran2021,Kagan2005,Delaney2004,9Barriers,15MDCobstacles,16Teamwork} \\
    RC2 & To provide quality control for the recommendations. & 9 & \cite{Sutcliffe2004,Wright2007,Rajasekaran2021,Kagan2005,Delaney2004,10vMDT,13MDCoppur,15MDCobstacles,16Teamwork}\\ 
    RC3 & To allow detailed and open discussion. & 11 & \cite{Sutcliffe2004,Wright2007,Rajasekaran2021,Kagan2005,Delaney2004,9Barriers,10vMDT,12EHR,13MDCoppur,16Teamwork,GermanStudy}\\
    RC4 & To allow knowledge transfer between experts. & 12 & \cite{Sutcliffe2004,Wright2007,Rajasekaran2021,Kagan2005,Delaney2004,8CommBreakdown,9Barriers,12EHR,13MDCoppur,15MDCobstacles,16Teamwork,GermanStudy} \\
   
    & \textbf{Protocol Oriented Requirements, $RP$} &  & \\
    RP1 & To enable communication of patient history. & 8 & \cite{Sutcliffe2004,Wright2007,Rajasekaran2021,Kagan2005,Macaskill2006,8CommBreakdown,16Teamwork,GermanStudy}\\
    RP2 & To allow communication of critical points. & 9 & \cite{Sutcliffe2004,Wright2007,Rajasekaran2021,GermanStudy,8CommBreakdown,12EHR,13MDCoppur,15MDCobstacles,16Teamwork}\\
    RP3 & To allow explanations and clarifications in the discussion. & 6 & \cite{Sutcliffe2004,Kagan2005,Delaney2004,8CommBreakdown,9Barriers,12EHR}\\
    RP4 & To provide mechanism for resolving disagreements. & 5 & \cite{Sutcliffe2004,Kagan2005,Delaney2004,9Barriers,16Teamwork}\\
    RP5 & To promote equal participation from all participants. & 6 &\cite{Sutcliffe2004,Wright2007,Kagan2005,Delaney2004,9Barriers,16Teamwork} \\
    RP6 & To allow equal distribution of illocutionary force among participants. & 4 & \cite{Sutcliffe2004,Delaney2004,9Barriers,16Teamwork}\\   
      
    & \textbf{Implementation Oriented Requirements, $RI$} &  & \\
    RI1 & To have a coordinator for the dialogue.& 9 & \cite{Wright2007,Macaskill2006,8CommBreakdown,9Barriers,10vMDT,12EHR,13MDCoppur,15MDCobstacles,16Teamwork}\\
    RI2 & To record the dialogue history and conclusion. & 6 & \cite{Wright2007,Macaskill2006,10vMDT,13MDCoppur,16Teamwork,GermanStudy}\\
    RI3 & To protect patient privacy. & 3 & \cite{Wright2007,Rajasekaran2021,10vMDT}\\ \bottomrule
  \end{tabular}
\end{table*}

The requirements were then categorised into four classes depending on the mechanism through which they can be satisfied.  Table \ref{tab:reqs} lists the requirements according to their proposed categorisation. Each row presents the requirement id, description, number of papers in the literature that mentioned this requirement and references to the corresponding works. Each of the categories and their corresponding requirements are described next. 

\subsection{Agent Oriented Requirements}
This category represents communication requirements that are directly related to the dialogue participant. This category has only one requirement, labelled as $RA1$. It reflects that communication among different experts fares better in cases where the participants show strong communication skills such as self-confidence, assertiveness, amiability and politeness. 
This is especially true where there is an organisational hierarchy amongst the participants. Consequently, a collaboration framework should ideally try to offset the communication weaknesses of the participants as much as possible.

\subsection{Cooperation Oriented Requirements}
 The cooperation oriented requirements, with identifiers RC1 - RC4, stress different aspects of cooperation during the collaborative dialogue. $RC1$ expresses cooperation as a general goal to be fulfilled during the dialogue. $RC2$ outlines the goal for the cooperation itself: to achieve quality control over the recommendations. Achieving this quality control through consensus requires open and frank discussion amongst the participants. This is expressed by $RC3$. Finally, a fundamental aspect of cooperation is the transfer of knowledge between the participants. This is captured by $RC4$.

\subsection{Protocol Oriented Requirements}

Protocol oriented requirements cover communication and logistic aspects that should be handled at the protocol definition level. Six such requirements were identified. These are given identifiers RP1 - RP6. RP1 ensures that patient history (or observations pertaining to the issue at hand in case of non-medical domains) is explicitly stated during the dialogue so that any faulty assumptions can be countered. RP2 brings critical concerns of the participants to the forefront of the collaboration process. By doing this, it ensures that the participants reflect on these issues. Explanations and clarifications can be useful tools for transfer of understanding amongst the participants. This can promote cooperation and help to align the knowledge and thinking of the participants. Hence, RP3  formalises this need and makes it part of the dialogue. Similarly, RP4 ensures that the protocol design incorporates a conflict resolution mechanism. Finally, RP5 and RP6 mitigate against possible power dynamics resulting from the organisational structure that might influence the dialogue participants. RP5 ensures that the  protocol design incorporates inclusiveness while RP6 incorporates equality into the design.
\subsection{Implementation Oriented Requirements}
 Implementation oriented requirements express logistic concerns that can only be addressed at the implementation level. There are three such requirements which are given identifiers from RI1 - RI3. Since a collaborative dialogue between more than two experts can entail administrative overhead, most studies \cite{16Teamwork,15MDCobstacles,13MDCoppur} found that having a designated role to oversee this greatly improves the collaboration process. Hence, this is captured by RI1. RI2 captures the necessity of recording the dialogue history so that it can be referred back to at a later time if required. Finally, since expert collaborations generally cover confidential topics and data, RI3 ensures that any confidential information exchanged during the collaboration is protected.

We consider all the requirements to form a core part of discussions although some seem to come up more frequently in literature as compared to others. For example, requirements RC4, RP2, RI1, RC36 and RC2 are more frequently mentioned while some others such as RP6 and RI3 are not. Nevertheless these represent fundamental aspects of these exchanges.

\section{A Formal Dialogue System for Expert Collaboration \label{game}}

This section formally presents the Experts' Dialogue Game (EDG) which embeds explanation-based illocutionary forces in an inquiry dialogue in order to emulate the inquisitive, explanatory and cooperative aspects of real-life consultations.  This is done so that the dialogue game can generate richer reasoning traces and meet the needs of successful collaboration amongst experts.  An \emph{Inquiry Dialogue} is defined in Walton and Krabbe's popular typology of dialogue types \cite{Walton1995} as a collaborative discussion amongst participants to find out the answer to one or more questions when none of them is presumed to know the correct answer beforehand. Later, Walton \cite{Walton2011} introduces an \emph{Explanatory Dialogue} as a discussion between two or more participants in order to bring about a transfer of understanding from one to another. In this case, the participants already agree on the topic but differ in their understanding of it. 

\begin{definition}
A \emph{Dialogue Game (DG)} is a tuple $DG = (X,L,Loc,R)$ where $X$ is the set of participating agents, $L$ is a logical language which represents the dialogue content, $Loc$ is the set of permitted locutions and $R={CM \cup CB \cup TM \cup CS \cup T \cup PL}$ represents the sets of rules for \emph{commencement}, \emph{combination}, \emph{termination}, \emph{commitment}, \emph{turn-taking} and \emph{politeness} respectively.
\end{definition}

Each of these elements is described next.

\subsection{Participants, X}
The game requires two or more participating agents, each representing an expert in some area, belonging to the set $X={1,2, \dots n}$ where $n$ is the total number of agents in the system. 

\subsection{Content Language, L}
Each agent  has its own private knowledge base, represented as $\Sigma_i$ where $i \in X$. The knowledge base is expressed in the content language $L$. Table \ref{KBAgents} shows the initial knowledge bases of the agents for the running example in the content language $L$. It is a first order logic language with the following components:
\begin{itemize}
    \item Let $H = \{h_1, h_2, \dots , h_p\}$ be the set of all possible observations recorded in a dataset where each $h_i$ represents a  feature. The value for each feature belongs to the set $V=\{v | v $ is a categorical or non-categorical value $\}$. For the running example, this would be patient history recorded in a dataset and represented as atomic predicates and terms in first order logic such as $age(48)$, $gender(female)$, $symptom(fatigue)$, $symptom(constipation)$, $increase(weight)$, $increase(sleep)$ and $skin(dry)$.
    \item Let $D=\{d_i, \dots , d_m\}$ be the set of all verdicts. For the running example, this would include $diagnosis(depression)$, $diagnosis(anaemia)$ and $diagnosis(hypothyroidism)$.    
    \item Let $E= \{e_i, \dots, e_w\}$ be the set of all evaluative measures that can be recommended for a particular case. For the running example, these could be the tests identified by the medical experts to get more data on the patient's condition such as blood complete picture $test(blood\_CP)$, $test(TSH)$., $test(T4)$ and $test(T3)$ levels.
    \item Let $T=\{t_1, \dots, t_y\}$ be the set of all possible remedial measures that can be taken. For the running example, this could be the drugs prescribed or recommendations for behaviour change for the patient such as $prescribe(idoine)$ and $walk\_steps(10000)$.
    \item Let $C=\{c_1, \dots, c_x\}$ be the set of all concerns/critical points that can be raised by an agent for a particular case. For the running example, this could be points of concern that the medical experts identify for a particular patient such as $family\_history(diagnosis(autoimmune\_disorder))$. Then $A=E \cup T$ will be the set of all recommendations that can be made for a single case and $O = D \cup A \cup C$ will be set of all verdicts, recommendations and concerns that can be discussed during the dialogue game. 
    \item Let $K=\{k_1, \dots , k_z \}$ be the set of all atomic facts in the domain knowledge such that $\{H,A,D\} \subset K$. For the running example, these would be $symptom(headache)$ and $symptom(backpain)$.   
    \item Let $F=\{f_1, \dots , f_v \}$ be the set of all inferences from elements of $H$, $K$ and $O$ that make up the domain knowledge. For the running example, this could be  $symptom(fatigue) \wedge increase(weight) \wedge increase(sleep) \wedge symptom(constipation)  \wedge skin(dry) \rightarrow diagnosis(hypothyroidism)$.
\end{itemize}

\begin{table*}
  \caption{Initial knowledge bases of agents $\alpha, \beta, \gamma$ for the example dialogue}
  \label{KBAgents}
  \begin{tabular}{p{0.30\textwidth}p{0.30\textwidth}p{0.30\textwidth}}\toprule
    \textit{$\Sigma_\alpha$} &\textit{$\Sigma_\beta$}  &\textit{$\Sigma_\gamma$}  \\ \midrule
	  $h_1 - h_7$ & & \\
    $d_1 - d_4$ & $d_1 - d_4$ & $d_1 - d_4$ \\
    $e_1 - e_4$ & $e_1 - e_4$ & $e_1 - e_4$ \\
    $k_1$, $k_2$ & $k_1$, $k_2$ & $k_1$, $k_2$ \\
    $c_1$, $r_1$, $r_2$ \\
    $f_1$,$f_5$, $f_6$, $f_8$, $f_{12}$ & $f_3$,$f_5$, $f_6$, $f_8$, $f_{12}$  & $f_2$, $f_4 - f_{12}$  \\ 
   \midrule
   \multicolumn{3}{l}{Key to Formulas in the Knowledge Base.} \\
   \midrule
   $h_1$ $age(48)$ \\ 
   $h_2$ $gender(female)$ \\
   $h_3$ $symptom(fatigue)$ \\
   $h_4$ $symptom(constipation)$ \\
   $h_5$ $increase(weight)$ \\
   $h_6$ $increase(sleep)$ \\
   $h_7$ $skin(dry)$ \\
   $r_1$ $walk$ \\
   $r_2$ $healthy\_diet$ \\
   
   $d_1$ $diagnosis(depression)$ \\
   $d_2$ $diagnosis(anaemia)$ \\
   $d_3$ $diagnosis(hypothyroidism)$ \\
   $d_4$ $diagnosis(hyperthyroidism)$ \\
   
   $e_1$ $test(TSH)$ \\
   $e_2$ $test(T4)$ \\
   $e_3$ $test(T3)$ \\
   $e_4$ $test(blood\_complete\_count)$ \\

   $k_1$ $symptom(headache)$ \\
   $k_2$ $symptom(backpain)$ \\
   
   $c_1$ $family\_history(diagnosis(autoimmune\_disorder))$\\   
    \multicolumn{3}{l}{$f_1$ $symptom(fatigue) \wedge increase(weight) \wedge increase(sleep) \rightarrow diagnosis(depression)$}\\
    \multicolumn{3}{l}{$f_2$ $symptom(headache) \wedge symptom(backpain)  \rightarrow diagnosis(depression)$}\\
    \multicolumn{3}{l}{$f_3$ $symptom(fatigue) \wedge increase(weight) \wedge increase(sleep) \wedge symptom(cold\_hands) \rightarrow diagnosis(anaemia)$}\\
    \multicolumn{3}{l}{$f_4$ $symptom(fatigue) \wedge increase(weight) \wedge increase(sleep) \wedge symptom(constipation)  \wedge skin(dry) \rightarrow diagnosis(hypothyroidism)$}\\
    \multicolumn{3}{l}{$f_5$ $confirm(diagnosis(anaemia)) \rightarrow test(blood\_complete\_picture) $}\\
    \multicolumn{3}{l}{$f_6$ $confirm(diagnosis(hypothyroidism)) \rightarrow test(TSH) \wedge test(T4) $}\\
    \multicolumn{3}{l}{$f_7$ $confirm(diagnosis(subclinical\_hypothyroidism)) \rightarrow test(TSH) \wedge test(T4) \wedge test(T3) $}\\
    \multicolumn{3}{l}{$f_8$ $TSH(high) \wedge T4(low) \rightarrow diagnosis(hypothyroidism) $}\\
   \multicolumn{3}{l}{$f_9$ $TSH(high) \wedge T4(normal) \wedge T3(normal) \rightarrow diagnosis(subclinical\_hypothyroidism) $}\\
    \multicolumn{3}{l}{$f_{10}$ $skin(dry) \wedge family\_history(diagnosis(autoimmune\_disorder)) \rightarrow diagnosis(hypothyroidism) $}\\
    \multicolumn{3}{l}{$f_{11}$ $\rightharpoondown symptom(fatigue)$ $ \wedge \rightharpoondown increase(weight)$ $\wedge \rightharpoondown increase(sleep)$ $\wedge \rightharpoondown symptom(constipation)$  $\wedge \rightharpoondown skin(dry) \rightarrow$}\\
    \multicolumn{3}{l}{\quad $diagnosis(subclinical\_hypothyroidism)$} 
    \\
      \multicolumn{3}{l}{$f_{12}$ $confirm(diagnosis(hyperthyroidism)) \rightarrow test(TSH) \wedge test(T4) \wedge test(T3) $}\\
   \bottomrule
  \end{tabular}
\end{table*}

\subsection{Locutions, Loc \label{loc}}
Each locution, represented by the letter $\tau$, is of the form $\tau = loc_{i}(p)$ where $loc$ defines the speech act performed, $i \in X$ is the agent uttering the locution and $p \in L$ represents the content of the locution except for the $\mathit{prompt}$ locution where $p = \tau^{'} \in Loc$ represents a previous locution. All participants are assumed to be the receivers of all the messages so the receiver id is not tracked.  The set of permitted locutions is given in column 2 of Table \ref{tab:protocol}. The locutions can be grouped into four disjoint subsets such that $Loc = L1 \cup L2 \cup L3 \cup L4$ which cover different aspects of the collaborative dialogue. Respective classes are indicated in column 1 of Table \ref{tab:protocol}. The characteristics of each subset are discussed next.

\paragraph{L1 (Informational Locutions)} There are five locutions in this subset: \emph{observation, verdict, advise, concern} and \emph{assert}. These are labelled from L1.1 - L1.5 respectively. In the commencement phase, L1.1 - L1.4 are used to set the context of the dialogue. In the progress stage, all five can be used to introduce new knowledge into the conversation. While L1.1 to L1.4 are used to introduce facts pertaining to a specified topic, L1.5 (assert) is used for introducing inference rules that relate the content of the first four locutions to each other. No distinction is made between strict and defeasible facts and rules. 

\paragraph{L2 (Requests)} We use a simplified version of the typology for different explanation requests (and replies) presented by \cite{Olena2014}. This is because while they meet the conversational needs for a specific scenario in the financial domain, our protocol targets a general consultation setting between experts without going into domain specific  details. Consequently, three types of requests are included. A request for explanation, represented by \emph{$wh\mhyphen explain$}, when the claim is agreed upon but one participant requires the other to provide more formal details or give an informal opinion. A request for justification, represented by \emph{$wh\mhyphen justify$}, when one participant needs the other to back up their standpoint. Finally, a clarification request, represented by \emph{$wh \mhyphen clarify$} when the participants agree about the claim but one of them has missing links in the reasoning process and so asks for this missing information. All three types of requests are assumed to be as generic as possible and cover not only the \emph{why} aspect but also the \emph{what}. Hence they are framed as \emph{$wh \mhyphen requests$}. These locutions embed explanatory illocutionary forces in the main inquiry dialogue, allowing for the generation of richer traces of the inquiry process. This is critical for making the result of the inquiry dialogue explainable.

\paragraph{L3 (Replies)} This subset has five locutions \emph{explain, justify, clarify, agree} and \emph{retract} which are given identifiers from L3.1 - L3.5 respectively. The first three are locutions for answering the corresponding \emph{wh-requests} from the L2 subset while the last two cover other possible answers such as \emph{agree} and \emph{retract}. The protocol assumes that the agents are always able to clarify and explain, but not always able to justify, in which case they retract.

\paragraph{L4 (Management Locutions)}
This subset defines a total of three locutions, which are given identifiers from L4.1 - L4.3. These are \emph{prompt, end} and \emph{pass}. These manage the dialogue in different ways. \emph{prompt} serves two purposes: it allows the speaker to indicate to the other participants that they are awaiting a response on a particular locution and it can also be used during the termination stage to justify why a participant has disagreed to end the dialogue. \emph{end} indicates an acknowledgement by a participant that they are satisfied with the dialogue outcome, thus, giving their consent to end the dialogue. If they have an outstanding issue, they can refuse to give their consent to end the dialogue. In this case, they are invited to justify this by using \emph{prompt} to let other participants know which of their statements have not received a response yet.  Finally, since the dialogue game allows participants to make multiple moves, \emph{pass} is used to manage turn-taking. Whenever a participant has finished whatever they wanted to say (they are allowed to  use multiple locutions) in their turn, they signal the end of their turn by using \emph{pass}. Thus, in the case of more than two agents, the protocol allows everyone to participate in the explanatory dialogue since the dialogue game allows using multiple responses to each locution (a detailed description is provided in Section \ref{rules} when the dialogue rules are introduced). This means that in response to locutions from subset L2, other agents who were not directly addressed in the preceding \emph{wh-request} can choose to participate in the information exchange by making an appropriate move.

\subsection{Rules, R \label{rules}}
The game has three stages: an opening stage governed by \emph{Commencement Rules}, a progress stage governed by \emph{Combination Rules} and a termination stage described by \emph{Termination Rules} \cite{Prakken2005Coherence}. Each of these are described next, followed by commitment, turn-taking and politeness rules.

\begin{table*}
  \caption{Combination rules for consultation between two expert agents $\alpha, \beta \in X$}
  \label{tab:protocol}
  \begin{tabular}{p{0.02\textwidth}p{0.29\textwidth}p{0.61\textwidth}}\toprule
    \textit{Id} &\textit{Locution} & \textit{Reply} \\ \midrule

	L1.1 & $\mathit{observation}_{a}(H_{i})$ where $H_{i} \subset H$ &  $\mathit{agree}_{b}(H_{i})$, $\mathit{observation}_{\beta}(H_{j})$, $\mathit{assert}_{\beta}(F_{k})$,  $\mathit{wh\mhyphen clarify}_{\beta}(H_{k})$ where $H_{i} \subset H$, $H_k \subset H_i$, $H_{j} \in H$ and $F_k \subset F$. $H_i, H_k$ represent observations exchanged so far about the case while $H_j$ represents any new facts related to the case and $O_{m} \subset O$ represent any inference rules that apply to the observation respectively.\\ 

	L1.2 & $\mathit{verdict}_{\alpha}(D_{i})$ where $D_i \subset D$ & $\mathit{agree}_{\beta}(D_{k})$, 
	    $\mathit{wh\mhyphen explain}_{\beta}(D_{k})$ , $\mathit{wh\mhyphen justify}_{\beta}(D_{k})$, 
	    $\mathit{verdict}_{\beta}(D_{j})$ where 
	    $D_k \subset D_i$ and  $D_i, D_{j} \subset D$.\\

	L1.3 & $\mathit{advise}_{\alpha}(A_i)$ where $A_i \subset A$ & $\mathit{agree}_{\beta}(A_k)$, 
	    $\mathit{wh\mhyphen explain}_{\beta}(A_k)$, $\mathit{wh\mhyphen justify}_{\beta}(A_k)$, $\mathit{wh\mhyphen clarify}_{\beta}(F_j)$, 
	    $\mathit{advise}_{\beta}(A_j)$ where 
	    $A_i, A_j \subset A$, $A_k \subset A_i$ and $F_j$ is property of $a_j \in A_i$.\\

	L1.4 & $\mathit{concern}_{\alpha}(C_i)$ where $C_i \subset C$ & $\mathit{agree}_{\beta}(C_k)$, $\mathit{wh\mhyphen justify}_{\beta}(C_k)$, $\mathit{wh\mhyphen explain}_{\beta}(C_k)$, 
	$\mathit{wh\mhyphen clarify}_{\beta}(f_i)$ 
	where 
	$C_i \subset A$, $C_k \subset C_i$ and $f_i$ is property of $c_i \in C_i$.\\

    L1.5 & $\mathit{assert}_{\alpha}(F_i )$ where $F_i \subset F$ & $\mathit{agree}_{\beta}(F_i)$,  $\mathit{assert}_{\beta}(F_j)$ where $j \neq i$ and $F_i, F_k \subset F$. \\

	L2.1 & $\mathit{wh\mhyphen explain}_{\alpha}(\theta)$ where $\theta \in H \cup O$ & $\mathit{explain}_{\beta}(\phi)$ where $\phi \in K \cup F$. \\

	L2.2 & $\mathit{wh\mhyphen justify}_{\alpha}(\theta)$ where $\theta \in O$ & $\mathit{justify}_{\beta}(\phi)$, $\mathit{retract}_{\beta}(\theta)$
    where $\theta \in O$ and $\phi \in F$.\\

	L2.3 & $\mathit{wh\mhyphen clarify}_{\alpha}(\theta)$ where  $\theta \in H \cup O$ & $clarify_{\beta}(\phi)$ 
    where  $\theta \in H \cup O$ and $\phi \in H \cup F$.\\

	L3.1 & $\mathit{explain}_{\alpha}(\theta)$ where $\theta \in K \cup F$ & $\mathit{agree}_{\beta}(\theta)$, $\mathit{assert}_{\beta}(\psi)$, $\mathit{wh\mhyphen clarify}_{\beta}(\theta_i)$, $\mathit{explain}_{\gamma}(\theta_k)$
	 where $\theta, \theta_k \in K \cup F$, $\psi \in F$, $\theta_i \in H \cup O $ such that $\theta_i$ is related to $\theta$ and $c \neq b$.\\

    L3.2 & $\mathit{justify}_{\alpha}(\theta)$ where $\theta \in F$ & $\mathit{agree}_{\beta}(\theta)$,  $\mathit{assert}_{\beta}(\psi)$,  $\mathit{wh\mhyphen explain}_{\beta}(\theta_i)$, $\mathit{wh\mhyphen clarify}_{\beta}(\theta_j)$, 
    $\mathit{justify}_{\gamma}(\theta_k)$
     where $\theta, \theta_k \psi \in F$ such that $\theta \neq \psi \neq \theta_k$ and $\theta_i, \theta_j \in H \cup O$ such that $\theta_i, \theta_j$ are related to $\theta$ and $\theta_i \neq \theta_j$.\\

	L3.3 & $\mathit{clarify}_{\alpha}(\theta)$ where $\theta \in H \cup F$ & $\mathit{agree}_{\beta}(\theta)$,  $\mathit{assert}_{\beta}(\psi)$, 
	 $\mathit{wh\mhyphen explain}_{\beta}(\theta_i)$,  $\mathit{wh\mhyphen justify}_{\beta}(\theta_j)$, 
	 $\mathit{clarify}_{\gamma}(\theta_k)$
	 where $\theta, \theta_k \in H \cup F$, $\theta \neq \theta_k$, $\psi \in F$, $\theta_i \in K \cup F$ and $\theta_j \in F$ such that $\theta_j, \theta_j$ are related to $\theta$, $\theta_i \neq \theta_j$.\\
	
	L3.4 & $\mathit{agree}_{\alpha}(\theta)$ where $\theta \in K \cup F \cup O \cup H$. & -\\

	L3.5 & $\mathit{retract}_{\alpha}(\theta)$  where $\theta \in O$. & -\\

	L4.1 & $\mathit{prompt}_{\alpha}(Loc_k)$ where $Loc_k \subset Loc$ is the set of locutions moved so far. & Any of the valid responses entailed by each element of $Loc_k$.\\

	L4.2 & $\mathit{end}_{\alpha}$ & $\mathit{end}_{\beta}$, $\mathit{prompt}_{\beta}(Loc_k)$  where $Loc_k \subset Loc$. \\ 
	L4.3 & $\mathit{pass}_{\alpha}$ & - \\ \bottomrule
  \end{tabular}
\end{table*}

\paragraph{Commencement Rules}
The topic of the dialogue game can be one or more subsets of $O$. The game always starts with the initiator agent presenting the facts of the case (observations), its own conclusions (verdicts), corresponding recommendations (advice) and any critical points (concerns) it deems important. Thus, the first turn is composed of the first four locutions from the locution subset L1. A move, represented by $\mu$, is a tuple  $\mu = <\tau, \tau^{'}>$ where $\tau$ is the new locution being introduced in the current turn while $\tau^{'}$ is a previously introduced locution. $\tau^{'}$ is null for all moves made in the commencement phase.  Formally,  entries L1.1 to L1.4 in the \emph{Locution} column of Table \ref{tab:protocol} formally present the four \emph{Informational} locutions that are part of the first turn. Strict ordering is enforced on the four locutions that make up the first move and is given by the sequence L1.1 to L1.4 in the table.

\paragraph{Combination Rules}

The protocol allows participants to start new threads in the conversation at any time. This is achieved by allowing one or more locutions in the same turn where each locution corresponds to a move. For a move which uses the $prompt$ locution, $\tau^{'}$ is null. Repeating the same move is not allowed, however, repeating the same locution in the same turn with different content is allowed. Hence, the next participant can respond to any number of locutions from the previous turns. In doing so, they have to specify the locution they are responding to ($\tau^{'}$) and pick any of the valid responses for that locution as defined in Table \ref{tab:protocol} where the \emph{Reply} column indicates possible reactions to each locution from the \emph{Locution} column. Subscripts $\alpha$ and $\beta$ identify the agent playing the move. 

\paragraph{Termination Rules}
The dialogue terminates when all the participants agree to end it. Any participant can start the process for getting consent from others to end the dialogue. They can do this by using the \emph{end} locution. This signals the start of the termination stage. Since the participants are assumed to be assertive and cooperative, this means that anything that the participants do not explicitly challenge is taken for granted as an agreement. Hence, when the dialogue ends, all the participants are assumed to have agreed on all the elements of set $O$ under discussion. However, each voting for termination may not always end in successful termination since any participant can refuse to give their consent. They are then invited to highlight any outstanding issues they have by using the \emph{prompt} locution as explained in Section \ref{loc}. If this happens, the dialogue moves back into the progress stage. Otherwise, they give their consent to end the dialogue (and to accept all the statements that went unchallenged by them) by using the \emph{end} locution. A move which uses the \emph{end} locution also has $\tau^{'}$ set to null. 

\paragraph{Commitment Rules\label{para:cr}}
Dialogue games generally require each participant have their commitments publicly available in the form of a \emph{Commitment Store}. The commitments are created as a result of particular speech acts and they ensure accountability for the participants. This is useful for making the dialogue coherent and productive. We follow Hamblin's notion of commitment stores as done by \cite{Amgoud2003Prop} where an agent's commitment store, $CS(i)$ for agent $i \in X$, corresponds to a publicly available subset of its original knowledge base $\Sigma_i$. Specifically the commitment store $CS^t(i)$ at any time interval $t$ for agent $i \in X$ contains elements of $H_i \cup O_i \cup F_i$, where $H_i \subset H$, $O_i \subset O$ and $F_i \subset F$. These represent the observations, verdicts, advice and concerns known by each agent at any time during the game such that $CS^{t+1}(i) = CS^{t}(i) + \{c|c$ \textit{is a new commitment} $\}$. In addition we introduce a multilateral agreement store for all agents, $AS(MAS) = \{c | \exists i,j \in X \mathit{\ such \ that\ } c \in CS(i) \cap CS(j) $ and $c \in K \cup F \cup O \cup H \}$. That is, it contains the observations, verdicts, advice and concerns that have had a multilateral agreement at any time during the dialogue. As in the case of \cite{Amgoud2003Prop}, the union of all individual commitment stores reflect the dialogue state any time whereas $AS(MAS)$ shows the global agreements rather than the information state of the dialogue. $AS(MAS)$ can be considered as a collective agreement store which provides a summary of the agreements during the dialogue to all participants. This is because of the assumption that the agents are assertive and therefore, commit to any statement that they do not explicitly challenge. Locution L3.4 ($\mathit{agree}$) helps to incrementally build up this summary, making it easier to synchronise the collective commitments of all agents. The mechanism of how this works is explained in the commitment rules that follow. Each rule describes the changes to $CS(i)$ for agent $i \in X$ and $AS(MAS)$ in reaction to each locution.

\begin{itemize}
    \item \emph{C1}: For locutions subset L1 and L3.1 to L3.3, the content of the locution is added only to the individual commitment store of the speaker, $CS(i)$.
    \item \emph{C2}: For locution L3.4, the content of the locution is added to the individual commitment store of the speaker agent and also to $AS(MAS)$. If the content of L3.4 is already added to $AS(MAS)$ as a result of a previous application of C2, it is not added again.
     \item \emph{C3}: For a locution $loc(arg) \in L2$ where $arg \in L$, no changes are made to the individual commitment store of the speaker. However, if $arg$ had been added to $AS(MAS)$ following C2 earlier in the dialogue, it is removed from $AS(MAS)$ since it only contains multilateral agreements rather than bilateral ones.
    \item \emph{C4}: For all locutions belonging to L4, no changes are made to neither the individual commitment store of the speaker nor $AS(MAS)$.
    \item \emph{C5}: For L3.5, which represents a \emph{retract}, the content of the locution is removed from the commitment store of the speaker and from $AS(MAS)$ if it has been added following rule C2. 
    \item \emph{C6}: After the dialogue terminates, the union of all individual commitment stores minus the conflicts is added to $AS(MAS)$ following the notion of implicit agreement to all unchallenged statements as described in \emph{Termination Rules}.
\end{itemize}

\begin{figure}
    \centering
    \includegraphics[width= 0.70\textwidth]{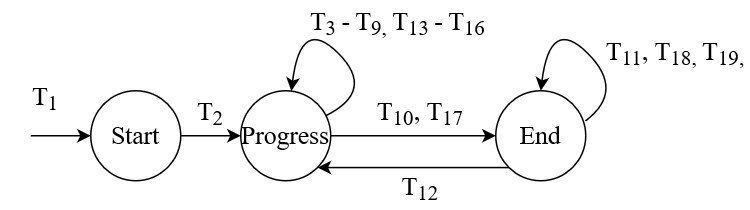}
    \caption{State transitions between commencement, progress and termination states for the example dialogue in Table \ref{tab:example}.}
    \label{fig:example-SM}
\end{figure}

Table \ref{tab:example} exemplifies the commencement, combination, commitment and termination rules for the running example. The first column of Table \ref{tab:example} indicates the turn identifier, the second column lists the identifier of the agent making the move, the third column identifies the locutions moved, the fourth column shows the changes in the commitment store of the speaker and the last column indicates the changes in the multilateral agreement store. In $T_1$, the first agent sets the context of the dialogue in accordance with the commencement rules. In $T_2$, agent $\beta$ asks $\alpha$ to justify their diagnosis of depression since its inference rule $f_3$ is in conflict with the diagnosis made by $\alpha$. In $T_4$, $\alpha$ provides this justification. However, in $T_3$, $\gamma$ provides its own diagnosis and challenges $\alpha$'s justification in $T_5$. Similarly, $\beta$ then asks $\gamma$ to justify their diagnosis of hypothyroidism in $T_6$ because it is in conflict with their inference rule $f_3$. $\gamma$'s justification is accepted by the other two agents in $T_8$ and $T_9$ respectively. In $T_13$, $\gamma$ asks $\alpha$ to explain why it recommends testing T3 levels. This is because it already has inference rule $f_12$ in its knowledge base but is confused with $\alpha$'s reasoning because of inference rules $f_7$ and $f_{11}$.  Fig. \ref{fig:example-SM} shows how the dialogue switches between the commencement, progress and termination states as a result of the different turns. Each node in the diagram represents a state with the state label given in the centre. Each arc represents a transition with the arc labels corresponding to the turn Ids in Table  \ref{tab:example}. 

 EDG promotes making justifications, explanations and clarifications explicit in the discussion by not allowing $assert$ in response to locutions $L1.2 - L1.4$. Table \ref{tab:extend} highlights how this can affect the dialogue. It shows how the running example in Table \ref{tab:introeg} would change after the twelfth move by $\alpha$ if $assert$ was allowed in response to locution $L1.3$. The first column indicates the identifiers of the alternative statements made by the experts. The identifiers for the alternative scenario are appended with a $\prime$ symbol to indicate that this is the alternative scenario. The second column shows the expert id and the statement they are making. Table \ref{tab:extend-CSMAS} shows the corresponding dialogue game. The columns are organised in the same way as for Table \ref{tab:example}. In this case, $\gamma$ does not ask $\alpha$ for an explanation, rather it provides it's own reasoning and the dialogue can close earlier. Hence, it is clear that EDG promotes justifications, explanations and clarifications at the expense of shorter discussion time.  

\begin{table*}
\begin{threeparttable}
  \caption{Example of a dialogue game between three medical expert agents, $\alpha, \beta, \gamma \in X$  \tnote{1}}
  \label{tab:example}
  \begin{tabular}{p{0.02\textwidth}p{0.05\textwidth}p{0.33\textwidth}p{0.25\textwidth}p{0.25\textwidth}}\toprule
     \textit{Id} &\textit{Speaker}  &\textit{Sequence of locutions} & \textit{ $CS(Speaker)$} & \textit{ $AS(MAS)$} \\ \midrule
    $T_1$ & $\alpha$ &
    $\mathit{observation}(h_1 - h_7)$ \newline  $\mathit{verdict}(d_1)$ \newline $\mathit{advise}(r_1, r_2)$ \newline $\mathit{concern}(c_1)$ \newline $\mathit{pass}$ &  $h_1 - h_7$ \newline $d_1$ \newline $r_1, r_2$  \newline  $c_1$ \\ 
	$T_2$ & $\beta$ & $<\mathit{wh\mhyphen justify}(d_1), \mathit{verdict}(d_1)>$ \newline $\mathit{pass}$ &  \\ 
	$T_3$ & $\gamma$ &
	$<\mathit{verdict}(d_3),\mathit{verdict}(d_1) >$ \newline  $\mathit{pass}$
 	&   $d_3$ \\ 
	$T_4$ & $\alpha$ & 
	$<\mathit{justify}(f_1), \mathit{wh\mhyphen justify}(d_1) >$ \newline $pass$
	&  $CS^{T_1}(\alpha) \cup f_1$ \\ 
	$T_5$ & $\gamma$ & $<\mathit{assert}(\rightharpoondown f_2), \mathit{justify}(f_1)>$ \newline $pass$ & $CS^{T_3}(\gamma) \cup$ $\rightharpoondown f_2$\\ 
	$T_6$ & $\beta$ & $<\mathit{wh\mhyphen justify}(d_3), \mathit{verdict}(d_3)>$ \newline  $\mathit{pass}$ &  \\

    $T_7$ & $\gamma$ & $<\mathit{justify}(h_7, c_1, f_{10}), \mathit{wh\mhyphen justify}(d_3)>$ \newline  $\mathit{pass}$ & $CS^{T5}(\gamma) \cup \{h_7, c_1, f_{10}\}$ \\

 	$T_8$ & $\alpha$& $<\mathit{agree}(h_7, c_1, f_{10}),\mathit{justify}(h_7, c_1, f_{10}) >$ \newline
    $<\mathit{advise}(e_1,e_2,e_3), \mathit{advise}(r_1, r_2)>$ \newline
    $\mathit{pass}$ & $CS^{T4}(\alpha) \cup$ \newline $\{h_7, c_1, f_{10},e_1,e_2,e_3\}$  & $\{h_7, c_1, f_{10}\}$\\

 	$T_9$ & $\beta$ & $<\mathit{agree}(e_1,e_2,e_3), \mathit{advise}(e_1,e_2,e_3)>$ \newline
    $<\mathit{advise}(e_4),\mathit{advise}(e_1,e_2,e_3)>$ \newline 
    $<\mathit{assert}(f_3), \mathit{justify}(h_7, c_1, f_{10})>$ \newline $\mathit{pass}$ &  $\{e_1,e_2,e_3,e_4,f_3\}$ & $AS^{T8}(MAS) \cup$ $\{e_1,e_2,e_3\}$\\
    
    $T_{10}$ & $\gamma$ & $<\mathit{agree}(e_1,e_2,e_4), \mathit{advise}(e_1,e_2,e_3)>$ \newline $\mathit{end}$ \newline $\mathit{pass}$ &  $CS^{T7}(\gamma) \cup \{e_1,e_2,e_4\}$ & $AS^{T9}(MAS) \cup$ $e_4$\\
 	$T_{11}$ & $\beta$ & $<\mathit{end},\mathit{end}>$ \newline  $\mathit{pass}$ & no change  \\
 	$T_{12}$ & $\alpha$ & $<\mathit{prompt}, \mathit{advise}(e_1,e_2,e_3)>$ \newline  $\mathit{pass}$ & no change \\
 	$T_{13}$ & $\gamma$ & $<\mathit{wh\mhyphen explain}(e_3), \mathit{advise}(e_1,e_2,e_3)>$ \newline $\mathit{pass}$ & no change  & $AS^{T10}(MAS)$ - $e_3$\\
 	$T_{14}$ & $\alpha$ & $<\mathit{explain}(f_{12}), \mathit{wh\mhyphen explain}(e_3)>$ \newline $\mathit{pass}$ 
  &  $CS^{T8}(\alpha) \cup$ $f_{12}$\\
  $T_{15}$ & $\beta$ & $<\mathit{agree}(f_{12}), \mathit{explain}(f_{12})>$ \newline  $\mathit{pass}$ 
  &  $CS^{T9}(\beta) \cup f_{12}$ & $AS^{T13}(MAS) \cup$ $f_{12}$\\
  $T_{16}$ & $\gamma$ & $<\mathit{assert}(f_4,f_7,f_{11}),\mathit{explain}(f_{12})>$ \newline  $\mathit{pass}$ & $CS^{T10}(\gamma) \cup \{f_7,f_{11}\}$  \\
  $T_{17}$ & $\alpha$ & $<\mathit{agree}(f_4,f_7,f_{11}), \mathit{assert}(f_4,f_7,f_{11})>$ \newline 
  $\mathit{end}$ \newline $\mathit{pass}$ 
  &  $CS^{T14}(\alpha) \cup$ $\{f_4,f_7,f_{11}\}$ & $AS^{T15}(MAS) \cup$ $\{f_4,f_7,f_{11}\}$\\
  $T_{18}$ & $\beta$ & $<\mathit{end},\mathit{end}>$ \newline  $\mathit{pass}$ 
  &  \\
  $T_{19}$ & $\gamma$ & $<\mathit{end},\mathit{end}>$ \newline  $\mathit{pass} $ 
  &  & $AS^{T17}(MAS) \cup$  $\{h_1, h_2, h_3, h_4, h_5, h_6\} \cup$ \newline $\{r_1, r_2, d_3\}$ \\
   \bottomrule
  \end{tabular}
         \begin{tablenotes}
       \item [1] Subscripts of locutions are not mentioned for clarity.
     \end{tablenotes}
  \end{threeparttable}
\end{table*}

\begin{table*} 
  \caption{Alternative ending for the running example. $\alpha, \beta, \gamma$}
  \label{tab:extend}
  \begin{tabular}{p{0.04\textwidth}p{0.90\textwidth}}\toprule
    \textit{Id} &\textit{Dialogue}  \\ \midrule
    $13^{\prime}$ & $\gamma$: I don't think it is necessary to test T3 at this stage since she is not asymptomatic. \\
    $14^{\prime}$ & $\alpha$: Okay. I think we can close the discussion now. \\
    $15^{\prime}$ & $\beta$: I agree. \\
    $16^{\prime}$ & $\gamma$: I agree. \\
   \\ \bottomrule
  \end{tabular}
\end{table*}

\begin{table*}
\begin{threeparttable}
  \caption{Dialogue game between three medical expert agents for the alternative ending in Table \ref{tab:extend} \tnote{2}}
  \label{tab:extend-CSMAS}
  \begin{tabular}{p{0.02\textwidth}p{0.05\textwidth}p{0.30\textwidth}p{0.25\textwidth}p{0.25\textwidth}}\toprule
     \textit{Id} &\textit{Speaker}  &\textit{Sequence of locutions} & \textit{ $CS(Speaker)$} & \textit{ $AS(MAS)$} \\ \midrule 
 	$T_{13^{\prime}}$ & $\gamma$ & $<\mathit{assert}(f_7, f_{11}), \mathit{advise}(e_1,e_2,e_3)>$ \newline  $\mathit{pass}$ & $CS^{T10}(\gamma) \cup \{f_7,f_{11}\}$ &\\
 	$T_{14^{\prime}}$ & $\alpha$ & $<\mathit{agree}(f_7,f_{11}),\mathit{assert}(f_7, f_{11})>$ \newline
  $\mathit{end}$ \newline $\mathit{pass}$ &  $CS^{T8}(\alpha) \cup \{f_7,f_{11}\}$ & $AS^{T10}(MAS) \cup$ $\{f_7,f_{11}\}$\\
  	$T_{15^{\prime}}$ & $\beta$ & $<\mathit{end},\mathit{end}>$ \newline  $\mathit{pass}$ &  no change & \\
   	$T_{16^{\prime}}$ & $\gamma$ & $<\mathit{end},\mathit{end}>$ \newline  $\mathit{pass}$ &  no change & $AS^{T14}(MAS) \cup$  $\{h_1, h_2, h_3, h_4, h_5, h_6\} \cup$ \newline $\{r_1, r_2, d_3\}$\\
   \bottomrule
  \end{tabular}
         \begin{tablenotes}
       \item [2] Subscripts of locutions are not mentioned for clarity.
     \end{tablenotes}
  \end{threeparttable}
\end{table*}

The collective agreement store serves as the output of the multi-agent system. It allows the most relevant knowledge for the decision making to be pooled together in a systematic way which is more computationally efficient than pooling all the knowledge bases of the agents. In the process, it also preserves the privacy of the agents since only publicly shared information is used. This approach allows for the building of a modular explainable multiagent system in which the multiagent decisions can be made independently of the human-machine interface. For example, it can be used to provide justified decisions made by expert agents to a user using another explanatory protocol for human-machine interaction such as the one proposed by Ilia et al. \cite{Ilia2023}. In this case, the collective agreement store can serve as the interface between the two modules of the explainable Artificial Intelligence (XAI) system.  

 EDG relies on the locutions $agree$ and $retract$ to incrementally synchronise agreements from the agents' individual commitment stores to the collective agreement store. However, if the agents fail to use these markers sufficiently, the burden of synchronising the agreement store will move to the end of the dialogue, stressing computational resources. Hence, the agents need to be made aware that explicit agreements will make the protocol more effective. 

Promoting elicitation of justifications, explanations and clarifications allows EDG to keep track of collective agreements and resolve discrepancies in the agreement store. An example of this can be seen by comparing the example in Table \ref{tab:example}  with the alternative scenario presented in Table \ref{tab:extend-CSMAS}. In the first case, the $wh\mhyphen request$ in turn $T_{13}$ results in removing the disputed recommendation of $e_3$ from the agreement store. In contrast, since $\gamma$ never makes their stance explicit in the scenario in Table \ref{tab:extend-CSMAS}, $e_3$ remains in the collective agreement store when the dialogue ends. So, the assumption that the agents are assertive is very important to ensure the success of EDG. An unassertive agent might end up being committed to beliefs that are not consistent with its knowledge base. 

\paragraph{Turn-taking Rules}
EDG identifies two roles for the participants, \emph{initiator} and \emph{participant}. However, the initiator role ends after the first turn, whereby everyone becomes a participant. The initiator provides sufficient context for the dialogue through the locutions in the first turn. The protocol enforces turns but no particular turn-order is enforced.  Each agent has to move at least one locution in response to the proponent's moved locutions. Since multiple locutions are allowed in each turn, each agent has to end his turn with the \emph{pass} locution to mark that he is finished.  

\paragraph{Politeness Rules} Structurally, dialogue games can allow participants to respond only once to each move \emph{(single-reply)} or offer several responses as well \emph{(multi-reply)}, to use only one locution in each move \emph{(single-move)} or more than one \emph{(multi-move)} and to transfer the turn as soon as some objective condition is met \emph{(immediate-reply)} or later \emph{(non-immediate-reply)} \cite{Prakken2005Coherence}. Based on these definitions, we consider EDG to be \emph{multi-reply}, \emph{multi-move} and \emph{non-immediate-reply}. A brief discussion justifying each of these properties follows next.

\paragraph{Multi-reply} The protocol achieves this in three ways. The first two enable this property for the respondent while the last one enables the speaker to proactively demand an additional response. For the respondent, it allows multiple arguments in one turn by not imposing any restrictions on the number of arguments included as content of each locution. For two, it allows respondents to come back to earlier choice points in the dialogue since it does not impose the restriction  on addressing the preceding move. So they can move several arguments referring to different previous moves if desired. For the speaker, it enables them to direct the conversation back to issues that were not addressed to their satisfaction using locution L4.1. 

\paragraph{Multi-move} The protocol does not limit the number of locutions that can be moved in one turn by each participant (see Section \ref{rules}). Hence, it is by construction \emph{multi-move}. 
\paragraph{Non-immediate-reply} Since the protocol does not enforce an external condition to shift the turn, it allows each agent to complete its move uninterrupted and proactively transfer the turn, it is then \emph{non-immediate-reply}.

All these properties make EDG very flexible and close to natural conversation. However, this flexibility can lead to dialogues that are incoherent or compromise the explanability and cooperative aspects of the dialogue. Hence, it calls for introducing the same mannerisms in place in natural conversations that act to counter these complications in real life conversation. So, EDG introduces two such mannerism into the dialogue as \emph{politeness rules}. It identifies two such rules to ensure dialogue progression and conflict resolution. The first is related to \emph{$wh \mhyphen request$}. Since the protocol does not force a participant to respond to only the previous move, the participants can ignore explanation, clarification and justification requests, defeating the explanatory objectives of the dialogue. In order to mitigate this, the first politeness rule requires that all \emph{Wh-Requests} must be responded to by the addressee first before they are allowed to respond to any other locution. Other participants who were not the direct addressee, can respond to a \emph{$wh \mhyphen request$} if they choose to do so. This allows the participants to collaboratively build explanations. The second rule concerns $\mathit{prompt}$. A $\mathit{prompt}$  serves as a reminder to the participants that this particular agent is awaiting a response for the prompted locution. The second rule gives the receivers of the $\mathit{prompt}$, the flexibility to choose to respond to it immediately or in a later turn. 

\subsection{Semantics}\label{para:semantics}

We take a protocol-oriented view of Agent Communication Language (ACL) semantics \cite{Pitt2000Comm, Prakken2005Coherence}. In this view, the semantics and use of utterances should be defined at the dialogue level rather that at the level of individual locutions \cite{Prakken2005Coherence}. Pitt and Mamdani \cite{Pitt2000Comm} distinguish between the \emph{content} and \emph{conversational} states of the dialogue. The former is dependent on the \emph{information state} of the agent. \emph{Information state} of an agent reflects its knowledge base. Semantics at this level define the change in the agent's information state. The latter is determined by the speech acts exchanged earlier in the dialogue, which are referred to as \emph{the conversation state}. The conversation state can be described by the set of possible responses for each speech act. Consequently, the commitment rules described in Section \ref{para:cr} form the \emph{content} level semantics for the protocol while the combination rules given in Table \ref{tab:protocol} define the \emph{conversational} semantics for the dialogue. Since the protocol treats the commitment store as a subset of the agent's knowledge base, the commitment rules express \emph{post-conditions} about the agent's information state as a result of the speech act. Next we describe the \emph{pre-conditions} for making the move.  

\paragraph{Pre-conditions for Managing Information State} 

\begin{itemize}
    \item \emph{P1}. For locutions subsets L1 and L3.1 to L3.3, there are no constraints on the content except that it should be relevant to the dialogue topic.
    \item \emph{P2}. For locution subsets L2, L3.4 and L4.1, the content of the locutions should already be part of the commitment stores of one of the agents.
    \item \emph{P3}. For L3.5, the content of the locution should belong to the commitment store of the speaker.
    \item \emph{P4}. For L4.2 and L4.3 no conditions apply as there is no content.
\end{itemize}

\paragraph{Pre-conditions for Managing Conversation State}

\begin{itemize}
    \item \emph{P5}. For locution subsets L1, L2, L3 and L4.1, those imposed by Table \ref{tab:protocol}.
    \item \emph{P6}. For L4.2, the agent finds no conflicts or objections in the information state of the dialogue as represented by its own commitment store.
    \item \emph{P7}. For L4.3, the agent making this move must have used at least one other valid locution before this one.
\end{itemize}

While pre-conditions can relate to both constraints on the agent's information state \cite{Amgoud2000, Amgoud2003Prop} or to constraints on the conversational state of the dialogue \cite{Walton2011}, here we specify pre-conditions to manage the information and conversational states of the dialogue itself. The limits introduced on the content of locutions in Table \ref{tab:protocol} also form part of the pre-conditions for managing the information state of the dialogue. 
We require that the agents maintain dialogue history and do not repeat a locution with the same content.

\section{A Platform for Expert Collaboration \label{platform}}
A prototype of EDG was implemented as a web application in order to evaluate it with human experts through a user study. The web application allows the participants to `chat' while enforcing EDG protocol. However, the participants do not need to remember the protocol, the web application enforces it for them. For each `message' in the chat, it shows the possible locutions from Table \ref{tab:protocol} that can be used in response as a drop down menu. The drop down menu item for each locution shows the corresponding natural language representation for a good user experience. For example, the locution $\mathit{wh\mhyphen justify}$ is listed as \emph{Can you justify}. Moreover, while EDG uses domain neutral terminology, the prototype implementation uses domain specific terminology for an intuitive user experience. For example, the locution $observation$ is shown as \emph{Patient history is} while the locution $verdict$ is represented as \emph{I diagnose}. However, these representations do not violate the dialogue protocol in any way. Rather they demonstrate how EDG can be applied in a domain specific context for a real world application.  The participant can select a locution to frame their response and type their text in the corresponding text field.  This section provides details on the implementation and the user study design. 

\subsection{Implementation}
\begin{figure}
    \centering
    \includegraphics[width= 0.70\textwidth]{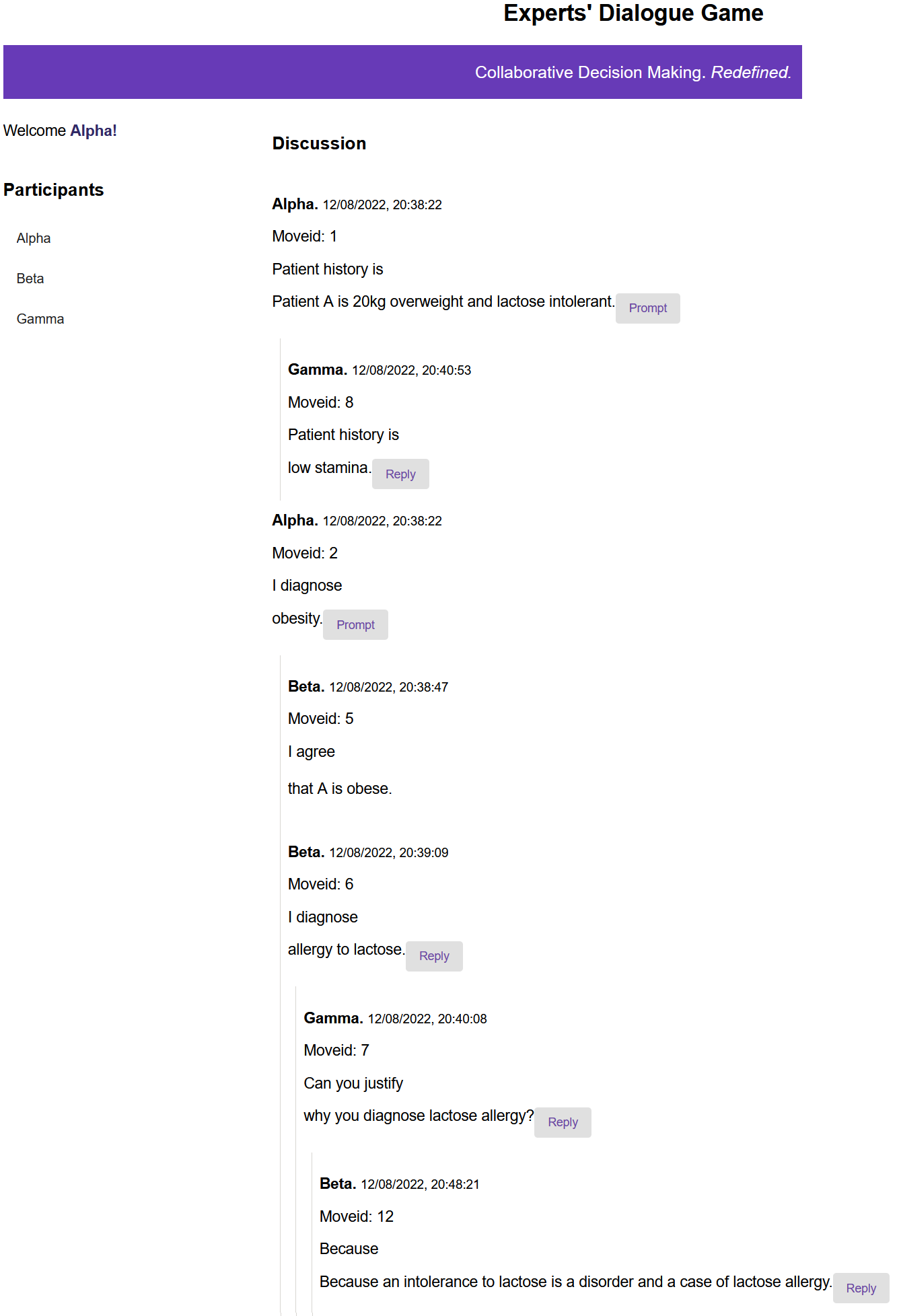}
    \caption{Screenshot of the web application implementing EDG.}
    \label{fig:screenshotEDG}
\end{figure}

EDG was implemented as a prototype full-stack web application that allows human participants to engage in discussion regarding the best diagnosis and treatment options for a patient. Fig. \ref{fig:screenshotEDG} shows a screenshot of graphical user interface of the web application with a hypothetical example. The application was implemented using JavaScript frameworks for client and server. The dialogue history is recorded in an SQLite database on the server. The server keeps track of the number of participants in a game and rotates the turns in a cyclic manner in the order in which the participants join the game session. The application enforces politeness rules by using highlights. It alerts the user who was the target of a \emph{wh-request} by highlighting the request in red and not allowing this user to play any other locution until all the \emph{wh-requests} to them have been discharged. Similarly, if a user plays the \emph{prompt} locution, the target locution is highlighted in blue on all participants' user interface to alert them on the request for response. However, they are not forced to respond to this alert. The web application was used to evaluate the usability of EDG through a user study. 

\subsection{User Study \label{ustudy}}
We conducted a user study to evaluate the \emph{usability} of the platform and its underlying protocol. According to the International Organization for Standardization (9241–11:2018), \emph{Usability} measures how \emph{effectively}, \emph{efficiently} and \emph{satisfactorily} a system, product or service can be used by the specified users for achieving their specified goals \cite{ISO}. We carried out \emph{formative usability testing}  with a total of six participants. \emph{Formative} usability testing is done during the development process with a relatively small number of participants to identify potential issues \cite{Barnum20219}. While traditional usability testing requires 30 to 50 test subjects, a type of \emph{formative} usability testing approach known as \emph{discount usability testing} has been shown to uncover $85\%$ of the issues for the task at hand with $5$ test subjects, with no significant subsequent  increase in the ratio of testing cost to benefits gained as the number of participants is increased \cite{Nielsen2000}.  Due to the highly specialised profile (i.e. medical experts) required for the test subjects in this user study, it was difficult to recruit test subjects. So discount testing was done with the minimum number of participants for the formative study. The goal of the study was to elicit user's perspectives on effectiveness, efficiency and satisfaction of the platform and the underlying protocol in meeting their professional communication needs.

\paragraph{Participants} 
The participants were final year medical students from a medical university in Barcelona, Spain. They were volunteers who responded to a call for participation after reading the advertised information sheet through their University's human resource department. 

\paragraph{Design}

 \begin{figure}
	\centering	\includegraphics[width=0.75\textwidth]{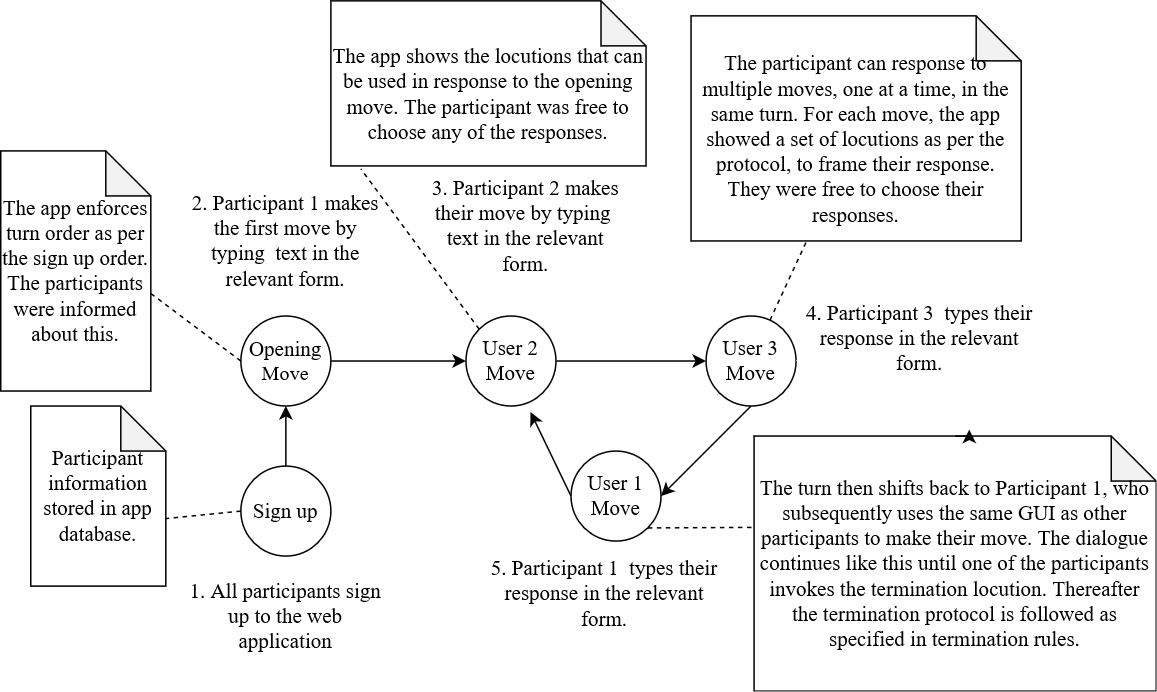}
	\caption{Workflow diagram of the user study design.}
	\label{fig:ustudy}
\end{figure}

For each session of the usability testing, participants were divided into groups of three. They were then tasked with collaboratively deciding on the best possible diagnosis and advice for an anonymous patient with a thyroid disorder. The patient data was taken from the publicly available thyroid dataset from the UCI ML repository (https://archive.ics.uci.edu/ml/datasets/Thyroid+Disease). Each participant was given an identical task description but the patient data was unequally distributed between the participants to see whether transfer of information would take place.  All sessions were conducted in a computer lab at a university where the participants had access to computers. All the participants in one session were in the same room at the same time. This was done to facilitate the administration of the study. However, participants were not allowed verbal or non-verbal (gestures such eye contact) communication with each other during the study in order to ensure that all communication took place through the web application. After the participants finished the task, semi-structured interviews were conducted to get their qualitative feedback on the application and the underlying protocol. The participants were given 30 minutes for the task and $11 - 20$ minutes for the post session interviews. All the participants who were in the same session were interviewed together as a group.  Figure \ref{fig:ustudy} shows the sequence of steps participants performed during the user study. First, all participants registered with the web application. The web application enforced the sign up order as the turn taking order in accordance with the turn taking rules. The first participant then made the opening move using the Graphical User Interface (GUI) of the web application. The turn was then passed onto the next participant in line. Each participant was notified by the web application when it was their turn. Each participant was free to choose which locutions they wanted to respond to and also which locution to use as their response. The online dialogue proceeded like this until one participant invoked the termination protocol by using the \emph{end} locution. Afterwards, the termination protocol was followed as described in termination rules in Section \ref{rules}. 

\paragraph{Goals}
The goal of the user study was to elicit perspectives of medical experts on aspects related to 
 effectiveness, efficiency, engagement and ease of learning for the discussion platform and the underlying protocol. These aspects also tie in to the requirement specification from Section \ref{Reqs}. However, they allow some additional usability considerations to be explicitly taken into account. Specifically, it aimed to answer the following questions for each of these aspects:

\begin{enumerate}
    \item   \textbf{Effective} 
    \begin{enumerate}
        \item Does the platform add value to professional discussions of medical experts?
        \item Did knowledge transfer take place as a result of the discussion? 
        \item Are participants satisfied with the explanations provided? 
        \item Are participants satisfied with the final decision and its justification? 
    \end{enumerate}
    
    \item \textbf{Efficient} 
    \begin{enumerate}
        \item Do participants have the moves they need at each step to express themselves? 
        \item Does the application impede the participants in some way during the dialogue? 
    \end{enumerate}
    
    \item \textbf{Engaging}  Do participants rate the experience as enjoyable? 
    
    \item \textbf{Easy to Learn} 
    \begin{enumerate}
        \item  Does the application and the protocol promote discussion? 
        \item Do participants find the classification of explanation requests useful or confusing ? 
    \end{enumerate}
\end{enumerate}

The next section provides details on the result of the user study. Additionally it provides a detailed evaluation of EDG according to the requirement specification of Section \ref{Reqs}.

\section{Evaluation Against Requirements for Expert Collaboration \label{eval}}
In this section we discuss how the protocol and its implementation measure up against the different types of requirements identified in Section \ref{Reqs}. We introduce three evaluation criteria, referred to as \emph{levels}, for this: \emph{dialogue}, \emph{system design} and \emph{user study}. Each of these is introduced next.

\paragraph{Dialogue} This level determines how well the dialogue rules presented in Section \ref{game} contribute towards satisfying each requirement. 
\paragraph{System Design} This level evaluates each requirement against the protocol implementation since some requirements can only be met at the implementation level.

\paragraph{User Study} This level validates the satisfaction of each requirement through the user study described in Section \ref{ustudy}.

Table \ref{tab:eval} presents a summary of the results. The first column shows the id of each requirement, columns second to fourth show whether the corresponding requirement is verified through the \emph{dialogue, system design} or \emph{user study}. Finally, the last column shows the representative quote from a participant in case the requirement satisfaction is verified through the user study.  The checkmark symbol in a table cell shows that the corresponding requirement is satisfied for the level indicated in the column header. The triangle symbol indicates that the corresponding requirement is satisfied indirectly by virtue of implementing the dialogue protocol. We distinguish it from the checkmark to show that the system design level does not make an active contribution to fulfilling the requirement for these cases. So, the evaluation for these requirements is superficial at this level. A blank value for the cell indicates that corresponding requirement is not satisfied by the level indicated in that column header.  Following paragraphs give a detailed discussion on each row of Table \ref{tab:eval}.

\begin{table*}[ht]
  \caption{Summary of requirements evaluation.}
  \label{tab:eval}
  \begin{tabular}{rc c cp{0.35\linewidth}}\toprule
    \textit{Id} & \multicolumn{3}{c}{\textit{Evaluated against}}  & \textit{Participant Quote}\\
    \cline{2-4}
    &  \textit{Dialogue } & \textit{System } & \textit{User} &
    \\ 
    &  \textit{} & \textit{ design} & \textit{ study} &
    \\ \midrule
    RA1 &  $\checkmark$ & $\triangle$ & $\checkmark$ &\textit{When answering the question if it felt natural to ask and be asked for clarifications and explanations etc}: ``Maybe sometimes there are people who will not tell you so directly (in real life) but the advantage of having this is that people will be more used to be asked for clarification and understand that it is the program that is predetermined.'' [Participant 5] \\ \midrule
    RC1 & $\checkmark$  & $\triangle$ & $\checkmark$  &  ``I think that it (the app) is more useful for medicals (doctors) to be in a team and to be more active if you .don't have 'I disagree' because (if) the other one doesn't give a correct explanation or contra answer and only puts `I disagree', you might take it personally, it may demotivate you." [Participant 4] \\ 
    RC2 &  $\checkmark$ &  &  & \\
    RC3 &  $\checkmark$ & $\triangle$  & $\checkmark$ & \textit{Answering the question if the app allowed them to have a productive discussion?} ``Yeah, it is interesting because we can have a quick chat to discuss case , it's a good option to discuss." [Participant 3]  \\
    RC4 &  $\checkmark$ & $\triangle$ & $\checkmark$ & ``A hundred per cent for me because I hadn't done the exam of Endocrinology so it was kind of fresh." [Participant 6]\\ \midrule
    RP1 &  $\checkmark$ & $\triangle$ &   & \\
    RP2 & $\checkmark$ & $\triangle$ &   & \\
    RP3 & $\checkmark$ & $\triangle$ &   & \\
    RP4 &  $\checkmark$ & $\triangle$ &   & \\
    RP5 &  $\checkmark$ & $\checkmark$ &   & \\
    RP6 &  $\checkmark$ &  &   & \\ \midrule
    RI1 &  & $\checkmark$ &   & \\
    RI2 &   & $\checkmark$ &   & \\
    RI3 &   & Partially &   &  \\
  \end{tabular}
\end{table*}

\subsection{Evaluation Against Agent Oriented Requirement}

\emph{RA1} is the only requirement that is directly derived from the participant's characteristics. Hence, it can be evaluated through the user study as well as with dialogue rules. By virtue of the fact that the dialogue rules already satisfy \emph{RA1}, the system design also satisfies the requirement inherently. We use the symbol $\triangle$ to indicate this. 

All locution groups \textbf{L1} to \textbf{L4} enable \emph{RA1} because presenting the set of possible options to speakers avoids failures from their possible lack of attention and assertiveness. It also enables richer dialogues by making the participants aware of the possible directions for branching out. This was also verified during the user study where the participants were of the opinion that the direct requests for explanations, clarifications and justifications allowed for more assertiveness and clearer communication. 
 
\subsection{Evaluation Against Cooperation Requirements}
 This section describes how the system and the dialogue game measures up against each of the cooperation oriented requirements on the three levels. 

At the dialogue level, \emph{RC1} is enabled through two mechanisms. The first is absence of any explicit locution to express disagreement such as `I disagree'. This allows the protocol to avoid deadlocks amongst participants. Secondly cooperation is enforced through implicit disagreements using locution subsets \textbf{L2} and \textbf{L3}. This was also verified during the user study where participants expressed the view that disallowing explicit disagreement was more useful in promoting cooperation and goodwill. \emph{RC1} is also part of the system design inherently but marked as $\triangle$ since it is superficially satisfied at the system design level.

\emph{RC2} is considered as a cooperation requirement since the dialogue protocol requires cooperation between the participants as a means for quality control. Hence, the protocol ensures \emph{RC2} because it requires at least one explicit agreement by another participant in order for the recommendation to be added to the collective commitment store. Even at that stage, they are open to non-monotonic debate. Hence at the end of the dialogue, only decisions that have survived the critical discussion are recommended. The embedding of explanatory illocutionary force, represented by locution \textbf{L2.2}, \textbf{L2.1} and \textbf{L2.3}, allows the participants to not only probe into each other's decisions but explanations as well. As in the case of decisions, only explanations which survive the critical debate make it into the collective commitment store. Although the dialogue rules provide a mechanism to ensure quality, the effectiveness of these quality control mechanism can only be verified in a deployment environment. Hence, the \emph{RC2} cannot be verified at the system design level so it is left as blank in the table. Similarly, since validating quality control in a deployment environment is a rigorous process and requires certified professionals, this requirement could not be validated within the scope of the current user study. Hence, it is left blank in the table.

 Since \emph{RC3} also requires a cooperative setting, it is listed as a cooperation requirement. The dialogue rules enforce \emph{RC3} by not forcing the participants to reply only to the preceding move, rather the rules allow participants the flexibility to reply to any number of previous locutions at any time. This allows any participant to either introduce new knowledge into the conversation by providing facts using locution rule subsets \textbf{L1} and \textbf{L3} or by asking questions using subset \textbf{L2}. Moreover, subsets \textbf{L2} and \textbf{L3} of locution rules allow the participants to probe into the statements of other participants, making detailed discussion possible. The dialogue is open since each participant has access to the dialogue state at all times. This was also verified during the user study where the participants felt that the discussion platform allowed them to have a \emph{productive} discussion. Since the system design also fulfils this requirement inherently by virtue of implementing the protocol, it is marked as $\triangle$ for system design.
 
\emph{RC4} is also dependent on cooperation of participants since only cooperation can ensure knowledge transfer. The dialogue game protocol enables this through locution classes \textbf{L1, L2 } and \textbf{L3}. Any member of these classes can be selected by the respondent to share their knowledge. This was verified during the user study because participants felt that they were able to gain new knowledge as a result of the information exchange. However, as before the system design satisfies this property inherently so it is marked as $\triangle$ under the system design column.  

\subsection{Evaluation Against Protocol Oriented Requirements}

 Next we evaluate the dialogue game against each of the six protocol oriented requirements in detail as summarised in Table \ref{tab:eval}.

 \emph{RP1} and \emph{RP2} are topic-based and lay down specific requirements for the inclusion of these topics. Their satisfaction can be verified through locution rules \textbf{L1.1} and \textbf{L1.4} which allow any participant to introduce patient history and critical points  into the dialogue at any time. 
 Moreover, there is no limitation on the amount of information transfer that can be done because the locutions can be repeated as many times as needed. The system design inherently satisfied these requirements so these are marked as $\triangle$ in Table \ref{tab:eval}. Verifying these requirements through the user study would be trivial, so these are marked as empty cells in Table \ref{tab:eval}. 
 
 The protocol incorporates \emph{RP3} through locution rules \textbf{L2.1, L2.3, L3.1} and \textbf{L3.3} which cover explanation and clarification requests as well as the corresponding responses. As before, since \emph{RP3} is inherently satisfied by system design, it is marked as $\triangle$ in Table \ref{tab:eval}. Similarly, it can only be verified superficially through the user study, so the corresponding cell in Table \ref{tab:eval} is left blank.
 
 The dialogue game provides mechanism to resolve conflicts through the introduction of locution rules \textbf{L2} and \textbf{L3}, thus satisfying \emph{RP4}. Through allowing for embedding of explanatory illocutionary forces within persuasive illocutionary forces and vice versa, disagreements are resolved indirectly by forcing the participants to spell out the nature of their disagreement rather than merely expressing it. For example, a disagreement due to need for evidence (represented by \textbf{L2.2}), as a result of missing link (\textbf{L2.3}) or a request for more information (\textbf{L2.1}). The participants can subsequently engage in a series of embedded  explanation dialogues until the issue is resolved to their satisfaction. Since the system design inherently satisfies the requirement, it is marked as $\triangle$ in Table \ref{tab:eval}. In the current user study, no conflicts appeared during the dialogues so this requirement could not be verified through the user study. Hence, it is left as blank in Table \ref{tab:eval}. 
 
 By enforcing turn-taking, the protocol ensures equal opportunity for getting input from all participants, satisfying \emph{RP5}. Since it is trivial to verify this requirement through the user study, it is marked as empty cell in Table \ref{tab:eval}. However, since the system design enforces a turn-taking mechanism, it satisfies \emph{RP5}.
 
 Finally, the protocol incorporates \emph{RP6} by giving the same validity to all moves by all participants. This requirement concerns the dialogue protocol definition so it does not make much sense to evaluate it through the user study or against system design. Hence, the corresponding cells are marked as empty in Table \ref{tab:eval}.

\subsection{Evaluation Against Implementation Oriented Requirements}
Next we discuss evaluation of each of the implementation oriented requirements according to the three evaluation criteria.

\emph{RI1} and \emph{RI2} require that the dialogue game be coordinated and the dialogue should be recorded. Both of these are satisfied at the system design level because the system acts as coordinator of the dialogue and does administrative book keeping tasks such as regulating turns, recording the dialogue and informing participants of the moves available at each time in the dialogue. Although the protocol description requires that these two requirements be met, it does not specifically provide a mechanism to enforce this. \emph{RI2} in particular, can only be enforced through system design. Hence, both these requirements are evaluated and verified at the system design level rather through dialogue rules or user study. So, the corresponding columns are marked as blank for the latter.

 The dialogue rules do not provide any mechanism to protect patient privacy as specified by \emph{RI3}. However, this requirement is partially satisfied through system design. This is because the platform limits the scope of information transfer to only the participants of the dialogue. Therefore, it protects patient privacy by design. Moreover, in the case of using anonymised data, it guarantees absolute privacy of the patient. However, the current prototype does not encrypt the information exchanged to secure it from malicious interference and leaves the anonymisation of the data exchanged to the participants' discretion. Hence, it only partially fulfils \emph{RI3}. Since it does not make much sense to evaluate it through the user study, the corresponding cell is marked as blank in Table \ref{tab:eval}.

\section{User Perspectives on Expert Collaboration System}\label{insights}
The post session interviews from the user study were recorded with the consent of participants, transcribed and the interview data was \emph{thematically analysed} to discover key insights. Thematic analysis is concerned with identifying patterns in qualitative data. The analysis can identify themes at the surface meaning level, referred to as \emph{semantic} level or go beyond what was said to discover underlying concepts, known as \emph{latent level} \cite{Maguire2017}.  Attention was paid to both semantic and latent meaning implied in the feedback. Specifically, the comments from the participants were organised into related concepts following the bottom-up organisation method of \emph{affinity matching}. Affinity matching is a bottom-up analysis technique for analysing data from a user study. In this case, relevant findings are grouped and the category labels are inferred from these groupings \cite{Barnum8}. The advantage as opposed to a top-down approach with pre-defined labels is that it keeps an open mind to what the data might reveal.  We identified six themes from the participants' feedback on various aspects of the dialogue game. Table \ref{tab:findings} shows the representative quotes form participants against each identified theme. Next, we describe the insights from users' perspectives for each theme.

\begin{table*}
  \caption{Representative quotes of participants for each emergent theme}
  \label{tab:findings}
  \begin{tabular}{rp{0.30\linewidth}p{0.55\linewidth}}\toprule
    \textit{S.No.} & \textit{Theme}  & \textit{Participant Quotes} \\ \midrule
   1 & Value in real-life use & ``Yeah it is a good tool because we have to talk with this specialist living in Madrid and it's a good chance to realise the questions or try to discuss on a case." [Participant 3] \newline ``...I had different values from my mate (for) TSH so I could receive more information from her and I think this is really useful because sometimes you can use the analytical values from another speciality where he has made the analytics and do interdisciplinary (exchange) between different specialities without repeating (the) exams because what they do is, that they already have the exam and when they go to a different doctor from another, they repeat the exam." [Participant 5]\\
   2 & Satisfaction with move options &  ``I think it is like ideal, it would be a good real scenario. It doesn't happen but I think it would be much easier like that." [Participant 3] \newline ``.. Sometimes I feel like we need an option that I am agree but like ..it's not completely white or black." [Participant 2]\\
   3 & Utility of different explanation requests & ``Maybe not so hard [as] justify or why you think about this or why you think about that. Not ..not,  we don't use that kind of hard expressions because justify is like WHY you are saying that." [Participant 2] \newline \textit{Answering why they don't consider justification request as rude?} ``Because maybe she knows what she wants to say but the way she explained to us is not as clear as she liked so." [Participant 4]\\
   4 & Effect of turn-taking & ``No, I think that it is good to give your turn for speaking because if not all the people will share information really quick and it will be difficult to extract important things ...umm..and it gives you time to think what you are gonna say and correct it if you have any error but maybe ...it will be good if you could ask for the turn to the program and you can answer in the order you have asked for your turn because maybe in one moment I didn't have an answer and it was my turn and the other person needed to wait for me but I think it's a really good ..[illegible]." [Participant 5] \newline  ``..The part in turns, is challenging because if somebody else is writing something and you see the message and you have an idea that you would like to comment (on) but you have to wait for another one, then it could , I don't know, slip your mind (or) whatever and then like (you would) not give your advice or recommendation that you should." [Participant 6]\\
   5 & User interface &  ``Maybe it will be easier if the values, number values of the diagnosis are shown at the right of the screen in a box so that you don't need to look for them in the dialogue and you don't lose information." [Participant 4] \newline ``I liked the discussion is divided by topics. So if you have the I diagnose part, then you can see like what everybody has said about diagnosis and it's not like one message about diagnosing, one message about exams, treatment all over, that get's like really messy. It was easier to read all about one part." [Participant 6]\\
   6 & User study design & ``Yeah if it's in a clinical setting I think [Participant 1] would have had all that data and we would only advise on the data. It felt kind of weird." [Participant 2[ \newline 
   ``I am not a hundred per cent sure about giving fragments of patient history to each of us because is like even with software and stuff, for patients, you have already all the information." [Participant 6]\\
  \end{tabular}
\end{table*}

\paragraph{Value in real-life use \label{us:value}}
The most common theme was the practical value of such a discussion system in the medical community. All participants in group 1 agreed that the platform would facilitate professional communication between medical experts, especially multidisciplinary communication and communication between public and private sectors. Participants agreed that the platform facilitates knowledge transfer and allows for a productive discussion. One participant was of the opinion that the platform helps clear communication by taking on the burden of \emph{politeness}, in that doctors in general do not ask each other for clarifications and justifications directly as it could be considered rude but since this is part of the platform's functionality, it no longer seems like a personal affront, rather just the way the platform works. This is a really significant comment since providing a means of getting around personal issues that hamper communication was one of the basic requirements the platform and the underlying protocol aimed to fulfil.

\paragraph{Satisfaction with move options \label{us:options}}
One of the most important practical aspects of a dialogue game for expert discussion is whether it allows the participants to express themselves completely. This was evaluated through taking participants' feedback on whether they found the dialogue locution sufficient for their discussion. All participants in group 1 expressed overall satisfaction with the request-response options that are part of the underlying protocol. They thought that it presented the \emph{ideal} scenario. 3 out of 6 participants thought that the response \emph{I agree} was not sufficient by itself because they would have liked to express partial agreement as well. For example, \emph{I agree but }. One participant was of the view that not having any \emph{I do not agree} option explicitly was good because doctors are in general always coming to a deadlock because of this so it was useful to ask for an explanation instead rather than expressing explicit disagreement. 

\paragraph{Utility of different explanation requests}
In order to evaluate whether the classification of different explanation requests is useful in a practical scenario, participants were asked whether they found the categorisation useful. 3 out of 6 participants in group 1 felt that all explanation requests were useful and they would use them in their communication. However, the other half argued that a justification request sounded really aggressive and unnatural and they would never use it in a real world scenario. They would prefer to use the explanation request instead. However, all participants agreed that they were not concerned about the subtle differences between explanation, clarification and justification requests as long as the intention of asking for an explanation was conveyed.

\paragraph{Effect of turn-taking}
Since traditional collaboration between experts is real-time, it was an open research question for protocol design whether it should allow synchronous or asynchronous communication. In order to meet requirements \emph{RP5} and \emph{RP6}, the protocol was designed to be synchronous and participants were asked to give their feedback based on their experience.
Two different perspectives emerged on the turn-taking mechanism implemented in the platform and the underlying protocol. 2 out of 6 participants in group 1 felt that the turns were good and helped to coordinate the discussion. 1 participant felt that it would be useful to be able to edit your response when it was not your turn because you might remember something after your turn had passed. 1 participant was of the view that ideas slipped your mind while waiting for your turn while another participant felt that the turn-taking put a lot of pressure on you to say something during your turn even when you did not think that you had anything to say.  They suggested that a dynamic turn-taking mechanism would be better to preserve coordination of the discussion and to deal with the last two problems. They thought that an option to queue for the turn token would be best. 

\paragraph{User interface}
Although the main aim for the user study was to evaluate the dialogue game protocol, it was anticipated that participants would end up evaluating the system at the user interface level. This was confirmed in the user study where the participants ended up offering several valuable suggestions for user interface improvements. All participants in group 1 thought that the clear separation of the discussion into history, diagnosis, advice and concerns was very useful and efficient. 3 out of 6 participants seemed happy with the user interface. 3 out of 6 participants agreed that having the history on a side panel would be more efficient because it would help reduce scrolling time to check on it. 1 participant also felt that displaying history in separate lines would also improve presentation. One participant felt that nesting the messages would make the presentation more clear and efficient. 

\paragraph{User study design}
One of the themes that emerged was that the participants were surprised with how the task in the user study was designed. Specifically, they found the distribution of patient history amongst the different participants unnatural compared to their experiences in the professional scenarios. All participants in group 1 found the division of patient history between the participants unnatural compared to the professional real life scenarios in which all the history is presented upfront. While this was done to check if transfer of information took place, they argued that knowledge transfer and information took place irrespective of this small test. So for the second group all the patient history was provided upfront to participant 1. 

\section{Conclusion and Future Work \label{Conc}}
This work envisioned a human-artificial agent hybrid collaborative recommendation system. As a first step towards this end, it presented a requirement specification for collaborative interactions between experts and an inquiry dialogue game grounded in the specification. The dialogue game allows multiple expert agents to collaborative on the best recommendations for a user. The game combines explanatory illocutionary forces in an inquiry dialogue. The motivation for doing this is to make the inquiry process and consequently, the output of the multiagent system explainable by generating richer traces for the reasoning process itself. The game presents an approach towards incorporating explainability within multiagent systems. This work also presented an evaluation of the dialogue game against the requirement specification through a user study. The user study was also significant in that it highlighted the real life utility for such dialogue platforms in the medical domain. Such platforms can enable clearer and systematic communication across multiple healthcare disciplines and sectors.  This work proposes to import  the methodology of software engineering into the area of formal dialogues. This methodology consists of the following steps: the collection of requirements for a dialogue game in a selected domain application; the design and implementation of a  protocol according to these requirements; and the evaluation of a dialogue system against the requirements.

The next step would be to implement and evaluate EDG for a multiagent system. This would involve evaluating formal properties of the system such as deadlocks, livelocks and termination guarantees.   One possible approach for investigating runtime termination guarantees for EDG is to explore multiagent frameworks that can provide this kind of guarantee through an appropriate moderator role. For example, the governor role in Electronic Institutions \cite{Ameli} can be extended to close termination property at runtime. Another research direction could be to close the protocol against disruption by non-cooperative and malicious agents. It might also be interesting to investigate whether the protocol can do away with turn-taking and synchronous communication since it may not scale well to a system with many participants.  While EDG presents which locutions can be used in response to others, it does not investigate how an agent can select the best locution in response to another based on its knowledge base. This is a very important aspect to implement the protocol in a multiagent system. Hence, an important future direction is to develop reasoning strategies for agents for participating in the EDG. One possible approach for doing this is to explore argumentation-based reasoning for EDG on the lines of \cite{WarrantInquiry2009}. Subsequently, the next step would be to adapt and implement the protocol in a human-artificial agent hybrid system. Other possible interesting directions to investigate include investigating how well the requirement specification presented in this work generalises to other domains such as engineering or aviation. On the flip side, it can also be interesting to evaluate how well existing dialogue game protocols \cite{WarrantInquiry2009,Luke2012,PrakenDD} conform to the requirement specification through a user study. Finally, the prototype of the platform developed as part of this work can be refined and released as open source software for facilitating communication between experts in the healthcare domain.

The politeness rules introduced here show that in order to make dialogue games more human-centred, we need to introduce the same societal machinery being employed in real life conversation in order to safeguard the integrity of the dialogue in a computational context. 


\section*{Acknowledgement}
The work reported in this paper has been supported in part by the European Union’s Horizon 2020 research and innovation programme under the Marie Skłodowska-Curie Grant Agreement No. 860621, in part by the project 2021 SGR 00754 of the Catalan Government, in part by the Polish National Science Centre, Poland (Chist-Era IV) under grant 2022/04/Y/ST6/00001, in part by POB CyberDS of Warsaw University of Technology within the Excellence Initiative: Research University (IDUB) programme under grant 1820/1/Z01/POB3/2021, and in part by VW foundation (VolkswagenStiftung) under grant 98 542.



\bibliographystyle{vancouver} 
\bibliography{refs}

\begin{thebibliography}{10}

\bibitem{Hennessy2011}
Hennessy CH, Walker A.
\newblock {Promoting multi-disciplinary and inter-disciplinary ageing research in the United Kingdom}.
\newblock Ageing and Society. 2011;31(1):52-69.
\newblock Available from: \url{https://www.cambridge.org/core/journals/ageing-and-society/article/abs/promoting-multidisciplinary-and-interdisciplinary-ageing-research-in-the-united-kingdom/044A8FFB174A3BF742EBA7F7DB3A4BD1}.

\bibitem{Healthcare}
Taberna M, Moncayo FG, Jan{\'{e}}-Salas E, Antonio M, Arribas L, Vilajosana E, et~al.. {The Multidisciplinary Team (MDT) Approach and Quality of Care}. Frontiers Media S.A.; 2020.
\newblock Available from: \url{/pmc/articles/PMC7100151/ /pmc/articles/PMC7100151/?report=abstract https://www.ncbi.nlm.nih.gov/pmc/articles/PMC7100151/}.

\bibitem{Naveed2018}
Naveed S, Donkers T, Ziegler J.
\newblock {Argumentation-based explanations in recommender systems: Conceptual framework and empirical results}.
\newblock In: UMAP 2018 - Adjunct Publication of the 26th Conference on User Modeling, Adaptation and Personalization. New York, NY, USA: Association for Computing Machinery, Inc; 2018. p. 293-8.
\newblock Available from: \url{https://dl.acm.org/doi/10.1145/3213586.3225240}.

\bibitem{10vMDT}
Munro AJ, Swartzman S.
\newblock {What is a virtual multidisciplinary team (vMDT)?}
\newblock British Journal of Cancer. 2013 jun;108(12):2433-41.
\newblock Available from: \url{https://pubmed.ncbi.nlm.nih.gov/23756866/}.

\bibitem{Walton2011}
Walton D.
\newblock {A dialogue system specification for explanation}.
\newblock Synthese. 2011 oct;182(3):349-74.
\newblock Available from: \url{https://link.springer.com/article/10.1007/s11229-010-9745-z}.

\bibitem{Arioua2017EXplain}
Arioua A, Buche P, Croitoru M.
\newblock {Explanatory dialogues with argumentative faculties over inconsistent knowledge bases}.
\newblock Expert Systems with Applications. 2017 sep;80:244-62.

\bibitem{Madumal2019}
Madumal P, Miller T, Sonenberg L, Vetere F.
\newblock {A Grounded Interaction Protocol for Explainable Artificial Intelligence}.
\newblock Proceedings of the International Joint Conference on Autonomous Agents and Multiagent Systems, AAMAS. 2019 mar;2:1033-41.
\newblock Available from: \url{http://arxiv.org/abs/1903.02409}.

\bibitem{Dennis2021}
Dennis LA, Oren N.
\newblock {Explaining BDI agent behaviour through dialogue}.
\newblock In: Proceedings of the International Joint Conference on Autonomous Agents and Multiagent Systems, AAMAS. vol.~1. International Foundation for Autonomous Agents and Multiagent Systems (IFAAMAS); 2021. p. 429-37.

\bibitem{Ilia2023}
Stepin I, Budzynska K, Catala A, Pereira-Fari{\~{n}}a M, Alonso-Moral JM.
\newblock {Information-seeking dialogue for explainable artificial intelligence: Modelling and analytics}.
\newblock Argument {\&} Computation. 2023;Preprint:1-59.

\bibitem{Hamblin1970}
Hamblin CL.
\newblock {Fallacies}.
\newblock 1st ed. London, UK: Methuen; 1970.

\bibitem{Walton1995}
Walton D, Krabbe ECW.
\newblock {Commitment in Dialogue: Basic Concepts of Interpersonal Reasoning}.
\newblock Albany NY: State University of New York Press; 1995.
\newblock Available from: \url{https://books.google.com.pk/books?id=6nU8TpVmW08C}.

\bibitem{Iyad2007}
Maudet N, Parsons S, Rahwan I.
\newblock {Argumentation in multi-agent systems: Context and recent developments}.
\newblock In: Lecture Notes in Computer Science (including subseries Lecture Notes in Artificial Intelligence and Lecture Notes in Bioinformatics). vol. 4766 LNAI. Springer Verlag; 2007. p. 1-16.

\bibitem{Searle1969}
Searle JR.
\newblock {Speech Acts}.
\newblock Cambridge University Press; 1969.
\newblock Available from: \url{https://www.cambridge.org/core/product/identifier/9781139173438/type/book}.

\bibitem{zora2020}
Alama J, Knoks A, Uckelman SL.
\newblock {Dialogue Games in Classical Logic}.
\newblock In: Giese M, Kuznets R, editors. TABLEAUX 2011: Workshops, Tutorials, and Short Papers, Technical Report IAM-11-002. Bern: Universit{\"{a}}t Bern; 2011. p. 82-6.
\newblock Available from: \url{https://doi.org/10.5167/uzh-190012}.

\bibitem{PrakkenLaw}
Prakken H.
\newblock {A formal model of adjudication dialogues}.
\newblock Artificial Intelligence and Law. 2008 sep;16(3):305-28.
\newblock Available from: \url{https://link.springer.com/article/10.1007/s10506-008-9066-4}.

\bibitem{purchaseNego2003}
McBurney P, {Van Eijk} RM, Parsons S, Amgoud L.
\newblock {A Dialogue Game Protocol for Agent Purchase Negotiations}.
\newblock Autonomous Agents and Multi-Agent Systems. 2003 nov;7(3):235-73.

\bibitem{Snaith2016}
Janier M, Snaith M, Budzynska K, Lawrence J, Reed C.
\newblock {A System for Dispute Mediation: The Mediation Dialogue Game}.
\newblock Frontiers in Artificial Intelligence and Application. 2016;287:351-8.
\newblock Available from: \url{https://doi.org/doi:10.3233/978-1-61499-686-6-351}.

\bibitem{Snaith2018}
Snaith M, {De Franco} D, Beinema T, {Op Den Akker} H, Pease A.
\newblock {A dialogue game for multi-party goal-setting in health coaching}.
\newblock In: Frontiers in Artificial Intelligence and Applications. vol. 305. IOS Press; 2018. p. 337-44.
\newblock Available from: \url{https://ebooks.iospress.nl/doi/10.3233/978-1-61499-906-5-337}.

\bibitem{Olena2014}
Budzynska K, Rocci A, Yaskorska O.
\newblock {Financial Dialogue Games: A Protocol for Earnings Conference Calls}.
\newblock In: Frontiers in Artificial Intelligence and Applications. vol. 266. IOS Press; 2014. p. 19-30.
\newblock Available from: \url{https://ebooks.iospress.nl/doi/10.3233/978-1-61499-436-7-19}.

\bibitem{Amgoud2000}
Amgoud L, Maudet N, Parsons S.
\newblock {Modelling dialogues using argumentation}.
\newblock In: Proceedings - 4th International Conference on MultiAgent Systems, ICMAS 2000. Institute of Electrical and Electronics Engineers Inc.; 2000. p. 31-8.

\bibitem{Walton2009}
Walton D.
\newblock {Argumentation Theory: A Very Short Introduction}.
\newblock In: Argumentation in Artificial Intelligence. Boston, MA: Springer US; 2009. p. 1-22.
\newblock Available from: \url{http://link.springer.com/10.1007/978-0-387-98197-0{\_}1}.

\bibitem{RatsmaPRakken2022}
Prakken H, Ratsma R.
\newblock {A top-level model of case-based argumentation for explanation: Formalisation and experiments}.
\newblock Argument {\&} Computation. 2022;13:159-94.

\bibitem{Bex2008}
Bex F, Prakken H.
\newblock {Investigating stories in a formal dialogue game}.
\newblock In: Frontiers in Artificial Intelligence and Applications. vol. 172. IOS Press; 2008. p. 73-84.

\bibitem{Black2009}
Black E, Atkinson K.
\newblock {Dialogues that account for different perspectives in collaborative argumentation}.
\newblock Proceedings of The 8th International Conference on Autonomous Agents and Multiagent Systems. 2009;2:867-74.
\newblock Available from: \url{papers2://publication/uuid/851E4870-2CCD-4C21-BBAD-819066F85D11}.

\bibitem{ArgInquiry2007}
Black E, Hunter A.
\newblock {A generative inquiry dialogue system}.
\newblock In: Proceedings of the International Conference on Autonomous Agents. New York, New York, USA: ACM Press; 2007. p. 1014-21.
\newblock Available from: \url{http://portal.acm.org/citation.cfm?doid=1329125.1329417}.

\bibitem{WarrantInquiry2009}
Black E, Hunter A.
\newblock {An inquiry dialogue system}.
\newblock Autonomous Agents and Multi-Agent Systems. 2009 oct;19(2):173-209.
\newblock Available from: \url{www.cossac.org}.

\bibitem{Huang1995}
Huang J, Jennings NR, Fox J.
\newblock {An agent architecture for distributed medical care}.
\newblock In: Lecture Notes in Computer Science (including subseries Lecture Notes in Artificial Intelligence and Lecture Notes in Bioinformatics). vol. 890. Springer Verlag; 1995. p. 219-32.
\newblock Available from: \url{https://link.springer.com/chapter/10.1007/3-540-58855-8{\_}14}.

\bibitem{Huang1994}
Huang J, Jennings NR, Fox J.
\newblock {Cooperation in Distributed Medical Care}.
\newblock 2nd Int Conf on Cooperative Information Systems (CoopIS-94). 1994:255-63.
\newblock Available from: \url{https://eprints.soton.ac.uk/252139/}.

\bibitem{Beveridge2006}
Beveridge M, Fox J.
\newblock {Automatic generation of spoken dialogue from medical plans and ontologies}.
\newblock Journal of Biomedical Informatics. 2006;39(5):482-99.

\bibitem{Vasileiou2023}
Vasileiou SL, Kumar A, Yeoh W, Son TC, Toni F.
\newblock {DR-HAI: Argumentation-based Dialectical Reconciliation in Human-AI Interactions}.
\newblock ArXiv. 2023;(2306.14694).
\newblock Available from: \url{http://arxiv.org/abs/2306.14694}.

\bibitem{Rago2023}
Rago A, Li H, Toni F.
\newblock {Interactive Explanations by Conflict Resolution via Argumentative Exchanges}.
\newblock arXiv. 2023;(2303.15022).
\newblock Available from: \url{http://arxiv.org/abs/2303.15022}.

\bibitem{Sassoon2019}
Sassoon I, K{\"{o}}kciyan N, Sklar E, Parsons S.
\newblock {Explainable argumentation for wellness consultation}.
\newblock In: Lecture Notes in Computer Science (including subseries Lecture Notes in Artificial Intelligence and Lecture Notes in Bioinformatics). vol. 11763 LNAI. Springer Verlag; 2019. p. 186-202.
\newblock Available from: \url{https://doi.org/10.1007/978-3-030-30391-4{\_}11}.

\bibitem{Sassoon2021}
Sassoon I, K{\"{o}}kciyan N, Modgil S, Parsons S.
\newblock {Argumentation schemes for clinical decision support}.
\newblock Argument and Computation. 2021;12(3):329-55.

\bibitem{Kokciyan2021}
Kokciyan N, Sassoon I, Sklar E, Modgil S, Parsons S.
\newblock {Applying Metalevel Argumentation Frameworks to Support Medical Decision Making}.
\newblock IEEE Intelligent Systems. 2021;36(2):64-71.

\bibitem{EQRbot}
Castagna F, Garton A, McBurney P, Parsons S, Sassoon I, Sklar EI.
\newblock {EQRbot: A chatbot delivering EQR argument-based explanations}.
\newblock Frontiers in Artificial Intelligence. 2023;6.
\newblock Available from: \url{/pmc/articles/PMC10076765/ /pmc/articles/PMC10076765/?report=abstract https://www.ncbi.nlm.nih.gov/pmc/articles/PMC10076765/}.

\bibitem{Shaheen2021}
Shaheen Qua, Toniolo A, Bowles JKF.
\newblock {Argumentation-Based Explanations of Multimorbidity Treatment Plans}.
\newblock In: Lecture Notes in Computer Science (including subseries Lecture Notes in Artificial Intelligence and Lecture Notes in Bioinformatics). vol. 12568 LNAI. Springer Science and Business Media Deutschland GmbH; 2021. p. 394-402.
\newblock Available from: \url{https://doi.org/10.1007/978-3-030-69322-0{\_}29}.

\bibitem{Tolchinsky2006}
Tolchinsky P, Cort{\'{e}}s U, Modgil S, Caballero F, L{\'{o}}pez-Navidad A. {Increasing human-organ transplant availability: Argumentation-based agent deliberation}. Institute of Electrical and Electronics Engineers Inc.; 2006.

\bibitem{Modgil2005Pnacho}
Modgil S, Tolchinsky P, Cort{\'{e}}s U.
\newblock {Towards formalising agent argumentation over the viability of human organs for transplantation}.
\newblock In: Lecture Notes in Computer Science (including subseries Lecture Notes in Artificial Intelligence and Lecture Notes in Bioinformatics). vol. 3789 LNAI. Springer, Berlin, Heidelberg; 2005. p. 928-38.
\newblock Available from: \url{https://link.springer.com/chapter/10.1007/11579427{\_}95}.

\bibitem{Tolchinsky2007}
Tolchinsky P, Atkinson K, Mcburney P, Modgil S, Cort{\'{e}}s U.
\newblock {Agents deliberating over action proposals using the ProCLAIM model}.
\newblock In: Lecture Notes in Computer Science (including subseries Lecture Notes in Artificial Intelligence and Lecture Notes in Bioinformatics). vol. 4696 LNAI. Springer Verlag; 2007. p. 32-41.
\newblock Available from: \url{https://link.springer.com/chapter/10.1007/978-3-540-75254-7{\_}4}.

\bibitem{Dung1995}
Dung PM.
\newblock {On the acceptability of arguments and its fundamental role in nonmonotonic reasoning, logic programming and n-person games}.
\newblock Artificial Intelligence. 1995 sep;77(2):321-57.
\newblock Available from: \url{http://linkinghub.elsevier.com/retrieve/pii/000437029400041X}.

\bibitem{Xiao2021}
Xiao L, Hu K, Fox J.
\newblock {A Group Decision Description Language and its Clinical Application}.
\newblock In: ACM International Conference Proceeding Series. Association for Computing Machinery; 2021. p. 160-7.
\newblock Available from: \url{https://dl.acm.org/doi/10.1145/3488838.3488866}.

\bibitem{Patkar2012}
Patkar V, Acosta D, Davidson T, Jones A, Fox J, Keshtgar M.
\newblock {Using computerised decision support to improve compliance of cancer multidisciplinary meetings with evidence-based guidance}.
\newblock BMJ Open. 2012;2(3):e000439.
\newblock Available from: \url{http://dx.doi.org/10.1136/bmjopen-2011-000439}.

\bibitem{ArgXAISurvey2021}
Vassiliades A, Bassiliades N, Patkos T. {Argumentation and explainable artificial intelligence: A survey}. Cambridge University Press; 2021.
\newblock Available from: \url{https://doi.org/10.1017/S0269888921000011}.

\bibitem{FToniKR12}
Craven R, Toni F, Cadar C, Hadad A, Williams M.
\newblock {Efficient Argumentation for Medical Decision-Making}.
\newblock In: Proceedings of the Thirteenth International Conference on Principles of Knowledge Representation and Reasoning. KR12. AAAI Press; 2012. p. 598-602.

\bibitem{FanToni2013}
Fan X, Craven R, Singer R, Toni F, Williams M.
\newblock {Assumption-based argumentation for decision-making with preferences: A medical case study}.
\newblock In: Lecture Notes in Computer Science (including subseries Lecture Notes in Artificial Intelligence and Lecture Notes in Bioinformatics). vol. 8143 LNAI. Springer, Berlin, Heidelberg; 2013. p. 374-90.
\newblock Available from: \url{https://link.springer.com/chapter/10.1007/978-3-642-40624-9{\_}23}.

\bibitem{Chaudhry2006}
Chaudhry B, Wang J, Wu S, Maglione M, Mojica W, Roth E, et~al.. {Systematic review: Impact of health information technology on quality, efficiency, and costs of medical care}. American College of Physicians; 2006.

\bibitem{Carayon2019}
Carayon P, Hoonakker P.
\newblock {Human Factors and Usability for Health Information Technology: Old and New Challenges}.
\newblock Yearbook of medical informatics. 2019;28(1):71-7.
\newblock Available from: \url{https://pubmed.ncbi.nlm.nih.gov/31419818/}.

\bibitem{Tang2018}
Tang T, Lim ME, Mansfield E, McLachlan A, Quan SD.
\newblock {Clinician user involvement in the real world: Designing an electronic tool to improve interprofessional communication and collaboration in a hospital setting}.
\newblock International Journal of Medical Informatics. 2018;110:90-7.

\bibitem{Tang2019}
Tang T, Heidebrecht C, Coburn A, Mansfield E, Roberto E, Lucez E, et~al.
\newblock {Using an electronic tool to improve teamwork and interprofessional communication to meet the needs of complex hospitalized patients: A mixed methods study}.
\newblock International Journal of Medical Informatics. 2019;127:35-42.

\bibitem{Tang2023}
Nie JX, Heidebrecht C, Zettler A, Pearce J, Cunha R, Quan S, et~al.
\newblock {The Perceived Ease of Use and Perceived Usefulness of a Web-Based Interprofessional Communication and Collaboration Platform in the Hospital Setting: Interview Study With Health Care Providers}.
\newblock JMIR Human Factors. 2023;10(1):e39051.
\newblock Available from: \url{https://humanfactors.jmir.org/2023/1/e39051}.

\bibitem{Kurahashi2018}
Kurahashi AM, Stinson JN, van Wyk M, Luca S, Jamieson T, Weinstein P, et~al.
\newblock {The perceived ease of use and usefulness of loop: evaluation and content analysis of a web-based clinical collaboration system}.
\newblock JMIR Human Factors. 2018;5(1).
\newblock Available from: \url{https://pubmed.ncbi.nlm.nih.gov/29317386/}.

\bibitem{Lin2020}
Lin HJ, Ko YL, Liu CF, Chen CJ, Lin JJ.
\newblock {Developing and evaluating a one-stop patient-centered interprofessional collaboration platform in taiwan}.
\newblock Healthcare (Switzerland). 2020;8(3).
\newblock Available from: \url{https://pubmed.ncbi.nlm.nih.gov/32751264/}.

\bibitem{MartinezGarcia2013a}
Mart{\'{i}}nez-Garc{\'{i}}a A, Moreno-Conde A, J{\'{o}}dar-S{\'{a}}nchez F, Leal S, Parra C.
\newblock {Sharing clinical decisions for multimorbidity case management using social network and open-source tools}.
\newblock Journal of Biomedical Informatics. 2013;46(6):977-84.

\bibitem{Ngo2020}
Ngo V, Matsumoto CG, Joseph JG, Bell JF, Bold RJ, Davis A, et~al.
\newblock {The personal health network mobile app for chemotherapy care coordination: Qualitative evaluation of a randomized clinical trial}.
\newblock JMIR mHealth and uHealth. 2020;8(5):e16527.
\newblock Available from: \url{https://mhealth.jmir.org/2020/5/e16527}.

\bibitem{Morse2021}
Morse RS, Lambden K, Quinn E, Ngoma T, Mushi B, Ho YX, et~al.
\newblock {A mobile app to improve symptom control and information exchange among specialists and local health workers treating Tanzanian cancer patients: Human-centered design approach}.
\newblock JMIR Cancer. 2021;7(1).
\newblock Available from: \url{/pmc/articles/PMC8088847/ /pmc/articles/PMC8088847/?report=abstract https://www.ncbi.nlm.nih.gov/pmc/articles/PMC8088847/}.

\bibitem{MobApps2020}
Zheng C, Chen X, Weng L, Guo L, Xu H, Lin M, et~al.. {Benefits of mobile apps for cancer pain management: Systematic review}. JMIR Publications Inc.; 2020.
\newblock Available from: \url{https://mhealth.jmir.org/2020/1/e17055}.

\bibitem{Wu2011}
Wu R, Rossos P, Quan S, Reeves S, Lo V, Wong B, et~al.
\newblock {An evaluation of the use of smartphones to communicate between clinicians: A mixed-methods study}.
\newblock Journal of Medical Internet Research. 2011;13(3).
\newblock Available from: \url{https://pubmed.ncbi.nlm.nih.gov/21875849/}.

\bibitem{Wright2007}
Wright FC, {De Vito} C, Langer B, Hunter A.
\newblock {Multidisciplinary cancer conferences: A systematic review and development of practice standards}.
\newblock European Journal of Cancer. 2007 apr;43(6):1002-10.

\bibitem{Gross1987}
Gross GE.
\newblock {The Role of the Tumor Board In a Community Hospital}.
\newblock CA: A Cancer Journal for Clinicians. 1987 mar;37(2):88-92.

\bibitem{Sutcliffe2004}
Sutcliffe KM, Lewton E, Rosenthal MM.
\newblock {Communication Failures: An Insidious Contributor to Medical Mishaps}.
\newblock Academic Medicine. 2004;79(2):186-94.
\newblock Available from: \url{https://pubmed.ncbi.nlm.nih.gov/14744724/}.

\bibitem{8CommBreakdown}
Vargas I, Garcia-Subirats I, Mogoll{\'{o}}n-P{\'{e}}rez AS, Ferreira-De-Medeiros-Mendes M, Eguiguren P, Cisneros AI, et~al.
\newblock {Understanding communication breakdown in the outpatient referral process in Latin America: A cross-sectional study on the use of clinical correspondence in public healthcare networks of six countries}.
\newblock Health Policy and Planning. 2018 may;33(4):494-504.
\newblock Available from: \url{https://pubmed.ncbi.nlm.nih.gov/29452401/}.

\bibitem{9Barriers}
Liu P, Lyndon A, Holl JL, Johnson J, Bilimoria KY, Stey AM.
\newblock {Barriers and facilitators to interdisciplinary communication during consultations: A qualitative study}.
\newblock BMJ Open. 2021 sep;11(9).
\newblock Available from: \url{https://pubmed.ncbi.nlm.nih.gov/34475150/}.

\bibitem{Delaney2004}
Delaney G, Jacob S, Iedema R, Winters M, Barton M.
\newblock {Comparison of face-to-face and videoconferenced multidisciplinary clinical meetings}.
\newblock In: Australasian Radiology. vol.~48. Wiley Press; 2004. p. 487-92.
\newblock Available from: \url{https://pubmed.ncbi.nlm.nih.gov/15601329/}.

\bibitem{GermanStudy}
Schellenberger B, Diekmann A, Heuser C, Gambashidze N, Ernstmann N, Ansmann L.
\newblock {Decision-making in multidisciplinary tumor boards in breast cancer care – an observational study}.
\newblock Journal of Multidisciplinary Healthcare. 2021;14:1275-84.
\newblock Available from: \url{/pmc/articles/PMC8179814/ /pmc/articles/PMC8179814/?report=abstract https://www.ncbi.nlm.nih.gov/pmc/articles/PMC8179814/}.

\bibitem{15MDCobstacles}
{Look Hong} NJ, Gagliardi AR, Bronskill SE, Paszat LF, Wright FC.
\newblock {Multidisciplinary cancer conferences: exploring obstacles and facilitators to their implementation.}
\newblock Journal of oncology practice. 2010 mar;6(2):61-8.
\newblock Available from: \url{http://www.ncbi.nlm.nih.gov/pubmed/20592777 http://www.pubmedcentral.nih.gov/articlerender.fcgi?artid=PMC2835483}.

\bibitem{16Teamwork}
Lamb BW, Sevdalis N, Arora S, Pinto A, Vincent C, Green JSA.
\newblock {Teamwork and team decision-making at multidisciplinary cancer conferences: Barriers, facilitators, and opportunities for improvement}.
\newblock World Journal of Surgery. 2011 sep;35(9):1970-6.
\newblock Available from: \url{https://pubmed.ncbi.nlm.nih.gov/21604049/}.

\bibitem{12EHR}
Quinn M, Forman J, Harrod M, Winter S, Fowler KE, Krein SL, et~al.
\newblock {Electronic health records, communication, and data sharing: Challenges and opportunities for improving the diagnostic process}.
\newblock Diagnosis. 2019 aug;6(3):241-8.
\newblock Available from: \url{https://pubmed.ncbi.nlm.nih.gov/30485175/}.

\bibitem{Macaskill2006}
Macaskill EJ, Thrush S, Walker EM, Dixon JM.
\newblock {Surgeons' views on multi-disciplinary breast meetings}.
\newblock European Journal of Cancer. 2006 may;42(7):905-8.

\bibitem{13MDCoppur}
Wright FC, Lookhong N, Urbach D, Davis D, McLeod RS, Gagliardi AR.
\newblock {Multidisciplinary cancer conferences: Identifying opportunities to promote implementation}.
\newblock Annals of Surgical Oncology. 2009 oct;16(10):2731-7.
\newblock Available from: \url{https://link.springer.com/article/10.1245/s10434-009-0639-6}.

\bibitem{Rajasekaran2021}
Rajasekaran RB, Whitwell D, Cosker TDA, Gibbons CLMH, Carr A.
\newblock {Will virtual multidisciplinary team meetings become the norm for musculoskeletal oncology care following the COVID-19 pandemic? - experience from a tertiary sarcoma centre}.
\newblock BMC Musculoskeletal Disorders. 2021 dec;22(1):18.
\newblock Available from: \url{https://pubmed.ncbi.nlm.nih.gov/33402136/}.

\bibitem{Kagan2005}
Kagan AR.
\newblock {The multidisciplinary clinic}.
\newblock International Journal of Radiation Oncology Biology Physics. 2005 mar;61(4):967-8.

\bibitem{Prakken2005Coherence}
Prakken H.
\newblock {Coherence and Flexibility in Dialogue Games for Argumentation}.
\newblock Journal of Logic and Computation. 2005 dec;15(6):1009-40.
\newblock Available from: \url{http://academic.oup.com/logcom/article/15/6/1009/1086845/Coherence-and-Flexibility-in-Dialogue-Games-for}.

\bibitem{Amgoud2003Prop}
Parsons S, Wooldkidge M, Amgoud L.
\newblock {Properties and complexity of some formal inter-agent dialogues}.
\newblock Journal of Logic and Computation. 2003 jun;13(3):347-76.

\bibitem{Pitt2000Comm}
Pitt J, Mamdani A.
\newblock {Communication Protocols in Multi-agent Systems: A Development Method and Reference Architecture}.
\newblock In: Issues in Agent Communication. Springer, Berlin, Heidelberg; 2000. p. 160-77.
\newblock Available from: \url{https://link.springer.com/chapter/10.1007/10722777{\_}11}.

\bibitem{ISO}
{ISO - ISO 9241-11:2018 - Ergonomics of human-system interaction — Part 11: Usability: Definitions and concepts}; 2018.
\newblock Available from: \url{https://www.iso.org/standard/63500.html}.

\bibitem{Barnum20219}
Barnum CM.
\newblock {1 - Establishing the essentials}.
\newblock In: Barnum CM, editor. Usability Testing Essentials (Second Edition). second edi ed. Morgan Kaufmann; 2021. p. 9-33.
\newblock Available from: \url{https://www.sciencedirect.com/science/article/pii/B9780128169421000010}.

\bibitem{Nielsen2000}
Nielsen J. {Why You Only Need to Test with 5 Users}; 2000.
\newblock Available from: \url{https://www.nngroup.com/articles/why-you-only-need-to-test-with-5-users/}.

\bibitem{Maguire2017}
Maguire M, Delahunt B.
\newblock {Doing a thematic analysis: A practical, step-by-step guide for learning and teaching scholars.}
\newblock All Ireland Journal of Higher Education. 2017;9(3).
\newblock Available from: \url{https://ojs.aishe.org/index.php/aishe-j/article/view/335}.

\bibitem{Barnum8}
Barnum CM.
\newblock 8.
\newblock In: Merken S, editor. {Analyzing the findings}. Elsevier; 2021. p. 287-319.

\bibitem{Ameli}
Esteva M, Rosell B, Rodr{\'{i}}guez-Aguilar JA, Arcos JL.
\newblock {AMELI: An agent-based middleware for electronic institutions}.
\newblock In: Proceedings of the Third International Joint Conference on Autonomous Agents and Multiagent Systems, AAMAS 2004. vol.~1; 2004. p. 236-43.

\bibitem{Luke2012}
Riley L, Atkinson K, Payne T, Black E.
\newblock {An implemented dialogue system for inquiry and persuasion}.
\newblock In: Lecture Notes in Computer Science (including subseries Lecture Notes in Artificial Intelligence and Lecture Notes in Bioinformatics). vol. 7132 LNAI. Springer, Berlin, Heidelberg; 2012. p. 67-84.
\newblock Available from: \url{https://link.springer.com/chapter/10.1007/978-3-642-29184-5{\_}5}.

\bibitem{PrakenDD}
Prakken H.
\newblock {Relating Protocols For Dynamic Dispute With Logics For Defeasible Argumentation}.
\newblock Synthese. 2001;127(1-2):187-219.

\end{thebibliography}

\end{document}


\begin{frontmatter}

\title{An Explanation-oriented Inquiry Dialogue Game for Expert Collaborative Recommendations
\\ \vspace{0.3cm} \textit{Supplementary Materials}
}
\runtitle{
An Explanation-oriented Inquiry Dialogue Game for Expert Collaborative Recommendations}

\end{frontmatter}

This document provides information on supplementary materials provided for the paper. These include the task descriptions provided to each participant, the post session interview questions, the transcripts of the two user study sessions and the screenshots of the web application.

\section{Task Descriptions}
The task descriptions given to each participant are provided in the document titled `Task Descriptions'. The participants could not see each others' description.

\section{Post Session Interview Questions}
The following list of questions was used to guide the interviews. However, not all questions were asked from the participants in each user study session, rather it was used as a guide to drive the discussion. 

\begin{enumerate}
    \item Did you feel unable to say what you wanted to at any stage because of the moves available to you? 
    \item What are some of the things you wanted to convey but could not? 
    
    \item Did you learn anything new as a result of the discussion? 
    \item Did you have any disagreements? If yes, were they resolved to your satisfaction? 
    \item Are you happy with the explanations you received during the discussion? 
    \item Are you satisfied with the final decision made as a result of the discussion? 
    \item How would you feel about using this kind of platform in your professional life?

    \item Did the discussion moves feel natural? 
    \item Did you feel some the moves available to you or made by the other player were incorrect or unnatural? If yes, were you able to rectify the situation? 

    \item Did the protocol hamper you in some way? 
    \item Did you feel in control during the dialogue or do you think the control was driven by the protocol? 
    \item Were there any surprises? Good or bad. \\satisfaction, open question
    \item Did you find the separation of different explanation types useful or confusing?
   
\end{enumerate}

\section{Transcripts}
The transcripts for the two user study sessions are provided as text files with titles `Transcript 1' and `Transcript 2'. In the transcripts, \emph{Researcher} refers to the interviewer, \emph{P1} is participant 1, \emph{P2} is participant 2 and \emph{P3} is participant 3. Since our ethics approval only allowed sharing them with the reviewers before publication, they need to be destroyed after publication. That was our commitment in the ethics approval. 

\section{Graphical User Interface of the Web Application}

This section shows screenshots of the web application. The screenshots show the dialogue between two hypothetical participants in order to illustrate different features of the application.

Figure \ref{fig2:signup} shows the sign up page for the participants. Two roles are offered, the Initiator and Participants, as described in the dialogue game. The person making the first move signs up as Initiator whereas everyone else signs up as Participant.

\begin{figure} [b]
    \centering
    \includegraphics[width= 0.70\textwidth]{figs2/1signup.png}
    \caption{Sign up page for participants.}
    \label{fig2:signup}
\end{figure}

\begin{figure}
    \centering
    \includegraphics[width= 0.70\textwidth]{figs2/2start.png}
    \caption{The form for making the first move used by Participant 1.}
    \label{fig2:open}
\end{figure}

\begin{figure}
    \centering
    \includegraphics[width= 0.70\textwidth]{figs2/2tprogress.png}
    \caption{The view of participant 1 after the opening moves.}
    \label{fig2:open-post}
\end{figure}

\begin{figure}
    \centering
    \includegraphics[width= 0.70\textwidth]{figs2/4turns2.png}
    \caption{The participants can click on `Finish Turn' button to transfer the turn to the next participant.}
    \label{fig2:finish-turn}
\end{figure}

\begin{figure}
    \centering
    \includegraphics[width= 0.70\textwidth]{figs2/3turns.png}
    \caption{The turn shifts to participant 2 after the first participants clicks on 'Finish Turn' button to end their turn.}
    \label{fig2:turnshift}
\end{figure}

\begin{figure}
    \centering
    \includegraphics[width= 0.70\textwidth]{figs2/5prompt-justify.png}
    \caption{Whenever a participants clicks on `Prompt', the app alerts the `prompted' user by highlighting the message they are supposed to respond to.}
    \label{fig2:prompt-alert}
\end{figure}

\begin{figure}
    \centering
    \includegraphics[width= 0.70\textwidth]{figs2/6post-prompt-justify.png}
    \caption{An answer by the prompted participant.}
    \label{fig2:prompt-ans}
\end{figure}

\begin{figure}
    \centering
    \includegraphics[width= 0.70\textwidth]{figs2/8end-request.png}
    \caption{Dialogue shown to all other participants whenever one of them clicks on `End' button to start the termination protocol.}
    \label{fig2:end}
\end{figure}